\documentclass[useAMS,usenatbib]{mn2e}

\usepackage{color}
\usepackage{graphics,graphicx}
\usepackage{psfig,subfigure}
 
\title[The non-dipolar fields of T Tauri stars]
      {The non-dipolar magnetic fields of accreting T Tauri stars}
\author[S. G. Gregory et al.]
{S. G. Gregory$^{1}$\thanks{E-mail: sg64@st-andrews.ac.uk}, 
 S. P. Matt$^{2}$, J.-F. Donati$^{3}$ and M. Jardine$^{1}$ \\
$^{1}$SUPA, School of Physics and Astronomy, University of St Andrews, North 
Haugh, St Andrews, Fife, KY16 9SS\\ 
$^{2}$Department of Astronomy, University of Virginia, P.O. Box 400325, Charlottesville, VA 22904-4325, USA\\
$^{3}$Laboratoire d'Astrophysique, Observatoire Midi-Pyr\'en\'ees, 
      14 Av. E. Belin, F-31400 Toulouse, France}

\begin{document}

\date{}

\pagerange{\pageref{firstpage}--\pageref{lastpage}} \pubyear{2008}

\maketitle

\label{firstpage}

\begin{abstract}
Models of magnetospheric accretion on to classical T Tauri stars often assume that stellar magnetic fields are 
simple dipoles.  Recently published surface magnetograms of BP~Tau and V2129~Oph have shown, however, 
that their fields are more complex.  The magnetic field of V2129~Oph was found to be predominantly 
octupolar.  For BP~Tau the magnetic energy was shared mainly between the dipole and octupole field components, 
with the dipole component being almost four times as strong as that of V2129~Oph.  From the published     
surface maps of the photospheric magnetic fields we extrapolate the coronal fields of both stars,
and compare the resulting field structures with that of a dipole.  We consider
different models where the disc is truncated at, or well-within, the Keplerian corotation radius.
We find that although the structure of the surface magnetic field is particularly complex for both stars, the 
geometry of the larger scale field, along which accretion is occurring, is somewhat simpler.  However, the 
larger scale field is distorted close to the star by the stronger field regions, with the net
effect being that the fractional open flux through the stellar surface is less  
than would be expected with a dipole magnetic field model.  Finally, we estimate the disc truncation radius,
assuming that this occurs where the magnetic torque from the stellar magnetosphere is comparable to the viscous
torque in the disc.
\end{abstract}



\section{Introduction}
Classical T Tauri stars (cTTSs) are young solar analogs which are accreting material from
circumstellar discs.   Many observations are consistent with the
scenario of the stellar field truncating the inner disc and channelling 
gas onto discrete regions of the stellar surface.   The shapes of near-IR spectral
energy distributions (e.g. \citealt{rob07}), and the kinematics of CO lines formed in the disc
\citep*{naj03}, are consistent with the disc having been truncated at a distance
of a few stellar radii.  Average surface fields of order a kG have been detected
on a number of cTTSs \citep{joh07}, which models suggest will be strong enough to disrupt the
inner disc \citep{kon91}.  The detection of inverse P-Cygni profiles, with
widths of several hundred kms$^{-1}$, can also be explained by material essentially
free-falling along the field lines of the stellar magnetosphere from the location of the inner
disc \citep{edw94}.  Furthermore, the excess continuum emission (veiling) in the optical and UV is 
likely to arise from shocks at the base of accretion funnels, with emission lines with high 
excitation potentials, e.g. HeI 5876{\AA }, forming mainly in such regions \citep{ber01}.      
  
The magnetic interaction between the stellar field and the disc may also have important 
consequences for the formation of planets.  Simulations by \citet{rom06}, and analytic work by \citet*{lin96} 
and \citet{fle08}, suggest that the inner disc hole, cleared by the star-disc interaction, may provide a natural barrier that 
decreases the rate of inward migration of forming planets.  There is also evidence that the large scale 
magnetosphere may directly disrupt the inner disc, producing warps which in some systems 
cross the observer's line-of-sight to the star (e.g. AA Tau, \citealt{bou07}).  Indeed 3D MHD
simulations have demonstrated that complicated warping effects in the disc truncation region
may be common for T Tauri stars \citep{rom03,rom04}.  

The star-disc interaction may also explain the slower rotation of cTTSs compared to the
typically older weak-line T Tauri stars, whose discs have largely dispersed (see e.g. the review
by \citealt{bou07IAU}).  Accretion of material from the inner disc would act to spin-up
the central star in the absence of some physical mechanism to remove angular momentum from the system.
Various magnetospheric accretion models have been developed to explain this mechanism, which differ in their 
assumed location of the inner disc, the details of how angular momentum is removed, and how the magnetic
field topology controls both accretion and outflows (\citealt{kon91};  
\citealt{col93}; \citealt{shu94}; \citealt*{fer00}; \citealt*{fer06}; \citealt*{kuk03}; \citealt{mat05a,mat05b,mat08a,mat08b}; 
\citealt*{lon05}; \citealt{bes08}).  Although observations indicate that the magnetic 
field topologies of cTTSs are complex (e.g. see the discussion in \citealt{gre06a}), the majority of
accretion models simply assume that the star has a strong dipolar magnetic field.  Recently
however a few studies have dropped this assumption, with \citet{gre05,gre06a} and \citet{jar06} being the first to 
consider how magnetic fields with a realistic degree of complexity would affect the accretion process. 
Other field geometries have since been considered by \citet*{lon07,lon08} who 
present the results of a 3D MHD simulation of
accretion to stars with composite fields that consist of dipole and quadrupole components tilted at various
angles relative to the stellar rotation axis.     

It is only recently, however, that instrumentation has advanced to the stage where it is possible
to map the magnetic topologies of classical T Tauri stars.  \citet{don07,don08a} have recently
published magnetic surface maps derived from Zeeman-Doppler imaging of the young solar-analogs 
V2129~Oph and BP~Tau, using data from the ESPaDOnS and NARVAL spectropolarimeters at 
the Canada-France-Hawai'i Telescope and T\'elescope Bernard Lyot respectively.    
Using the pre-main sequence evolutionary model of \citet*{sie00}, \citet{don07} determined 
that V2129~Oph, a high-mass T Tauri star of $1.35\,{\rm M}_{\odot}$, had developed a radiative
core.  In contrast, BP~Tau, at $0.7\,{\rm M}_{\odot}$, should be completely convective 
\citep{don08a}.  As the internal structure, and therefore the magnetic field generation process, 
is different in both V2129~Oph and BP~Tau, it is of particular interest to compare the 
structure and properties of their magnetic fields, as well as comparing both to a dipole. 

In \S2 we summarise the stellar parameters and magnetic field measurements of BP~Tau and V2129~Oph.  
In \S3 we describe how three-dimensional coronal fields can be extrapolated from Zeeman-Doppler
maps of photospheric magnetic fields, and compare the resulting field topologies with a large
scale dipole.  In \S4 and \S5 we discuss the structure of the open stellar, and accreting, field,
while \S6 contains our conclusions.


\section{The stars BP~Tau and V2129~Oph}
BP~Tau is one of the best studied classical T Tauri stars, with an abundance
of observational data across most wavebands.  In contrast to this, V2129~Oph is less 
well studied, despite being the brightest T Tauri star in the $\rho$-Oph star forming cloud.

\subsection{Stellar parameters}
For BP~Tau we adopt the same stellar parameters
as used by \citet{don08a} namely a mass of $M_{\ast}=0.7\,{\rm M}_{\odot}$, a radius 
of $R_{\ast}=1.95\,{\rm R}_{\odot}$ and a rotation period of $P_{rot}=7.6\,{\rm d}$. 
This implies an equatorial corotation radius,
\begin{equation} 
R_{co}=\left (\frac{GM_{\ast}P_{rot}^2}{4\pi^2}\right )^{1/3},
\end{equation}
of $7.4\,R_{\ast}$.  We assume a mass accretion rate of 
$2.88\times 10^{-8}\,{\rm M}_{\odot}{\rm yr}^{-1}$ \citep{gul98}.    

For V2129~Oph we adopt the same stellar parameters as used by \citet{don07} and \citet*{jar08}, 
namely a mass of $M_{\ast}=1.35\,{\rm M}_{\odot}$, a radius of $R_{\ast}=2.4\,{\rm R}_{\odot}$ and 
a rotation period of $P_{rot}=6.53\,{\rm d}$.  This implies 
an equatorial corotation radius of $R_{co}=6.7\,R_{\ast}$.  There are few estimates of the 
mass accretion rate on to V2129~Oph available in the literature.  \citet{eis05}, who refer
to V2129~Oph as AS~207A, estimate an accretion rate of $3.2\times 10^{-8}\,{\rm M}_{\odot}{\rm yr}^{-1}$,   
whereas \citet{don07} calculate a value of $4\times 10^{-9}\,{\rm M}_{\odot}{\rm yr}^{-1}$ using the 
empirical relationship between accretion rate and the flux in the CaII 8662{\AA} line presented by 
\citet*{moh05}.  We therefore assume that the disc mass accretion rate of V2129~Oph is 
$\dot{M} = 10^{-8}\,{\rm M}_{\odot}{\rm yr}^{-1}$, a compromise between both observationally 
inferred values.                
  

\begin{figure*}
        \def\subfigtopskip{4pt}
        \def\subfigbottomskip{4pt}
        \def\subfigcapskip{2pt}
        \centering
        \begin{tabular}{ccc}
                \subfigure{
                        \label{surface_bptau}
                        \psfig{figure=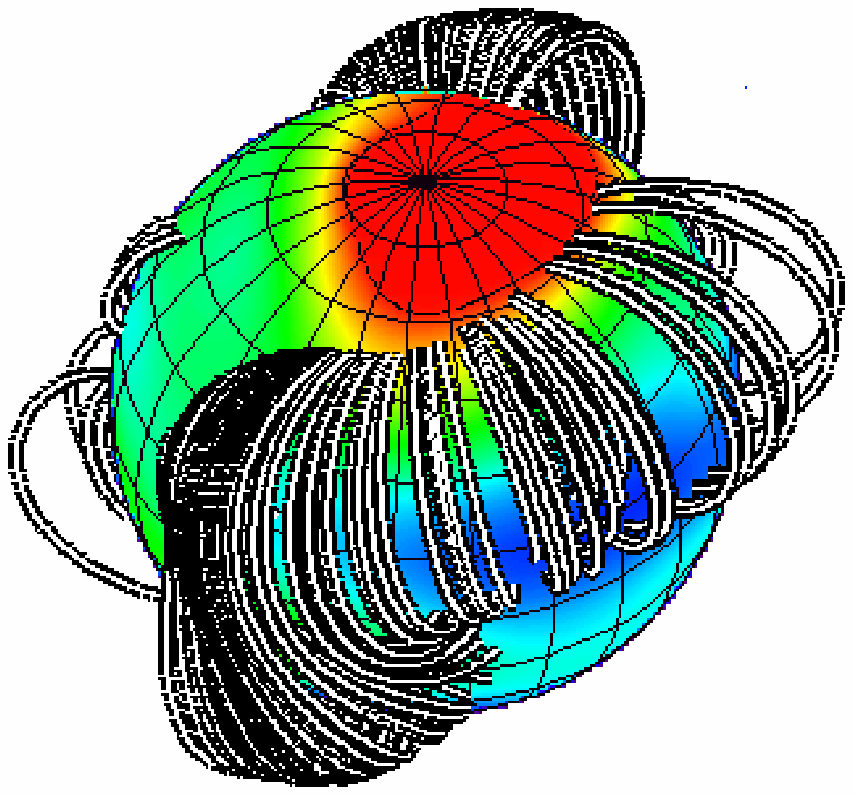,width=45mm}
                        } &
                \subfigure{
                        \label{accn_bptau}
                        \psfig{figure=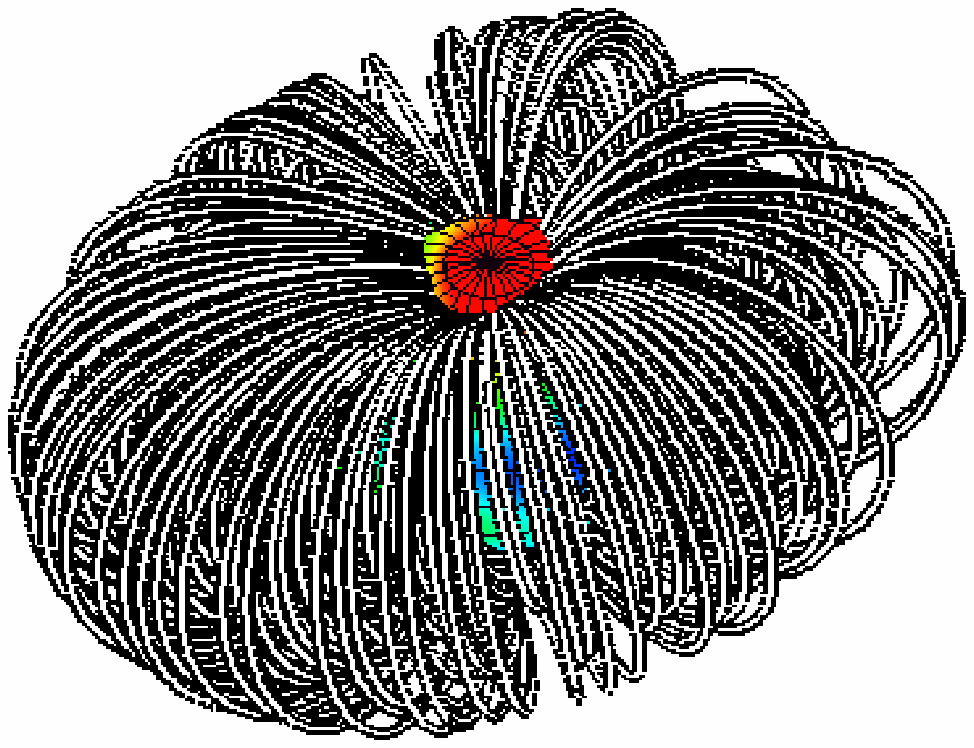,width=45mm}
                        } &
                \subfigure{
                        \label{open_bptau}
                        \psfig{figure=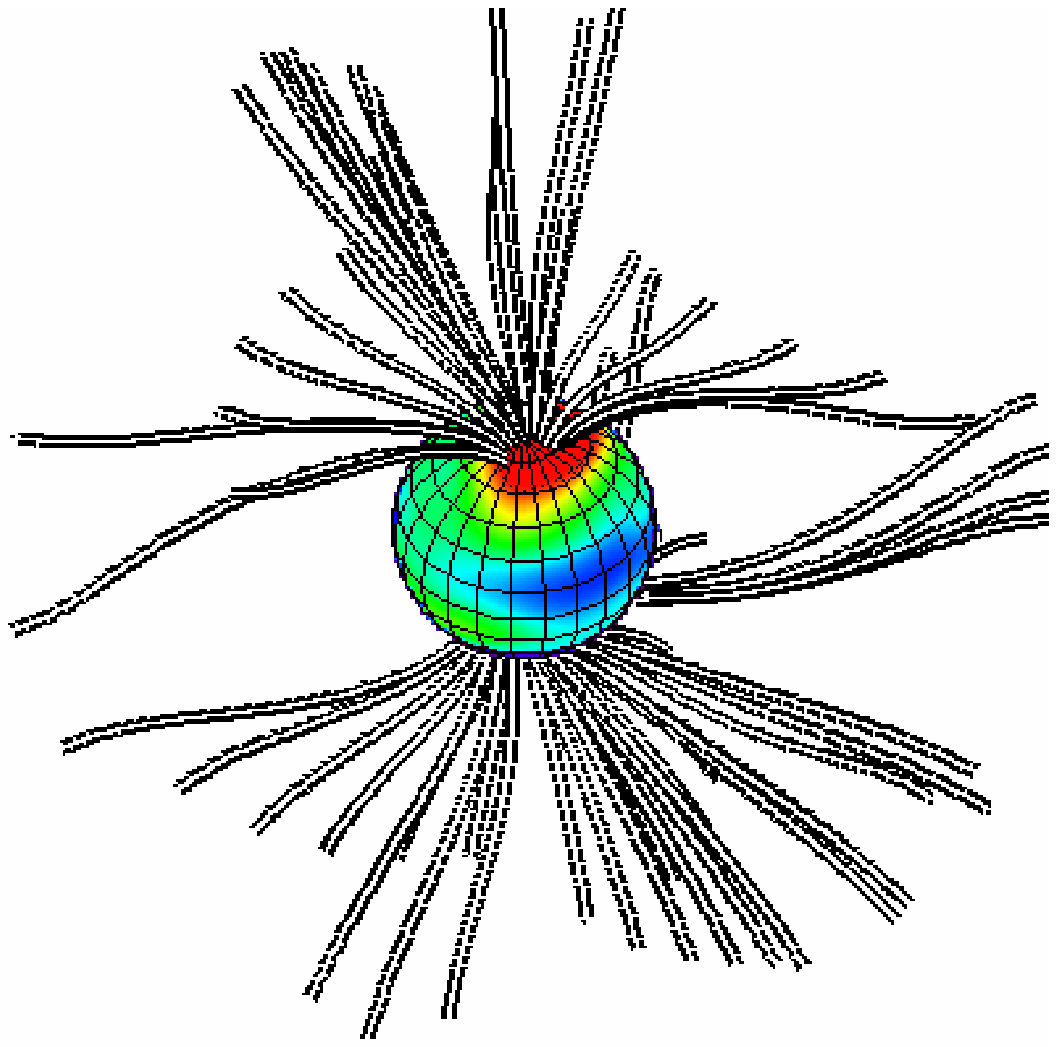,width=45mm}
                        } \\ 
                \subfigure{
                        \label{surface_v2129oph}
                        \psfig{figure=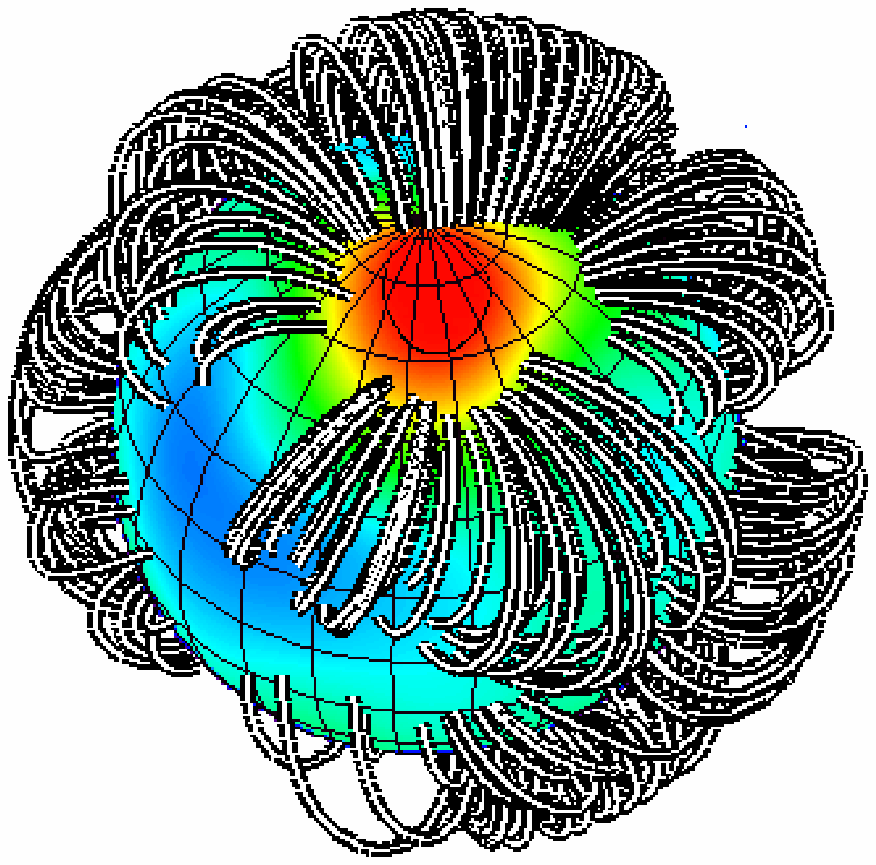,width=45mm}
                        } &
                \subfigure{
                        \label{accn_v2129oph}
                        \psfig{figure=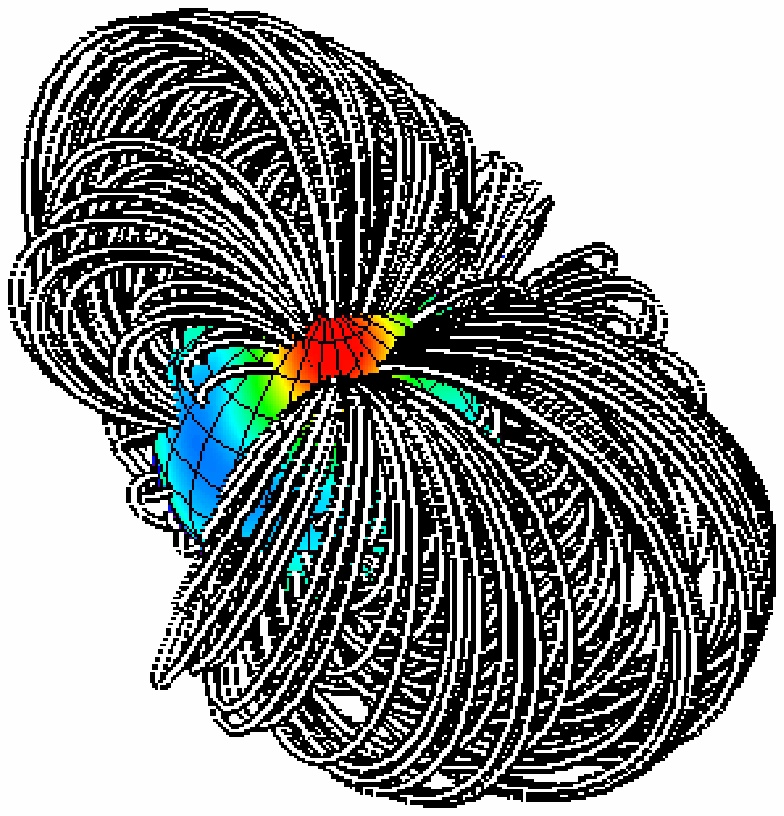,width=45mm}
                        } &
                \subfigure{
                        \label{open_v2129oph}
                        \psfig{figure=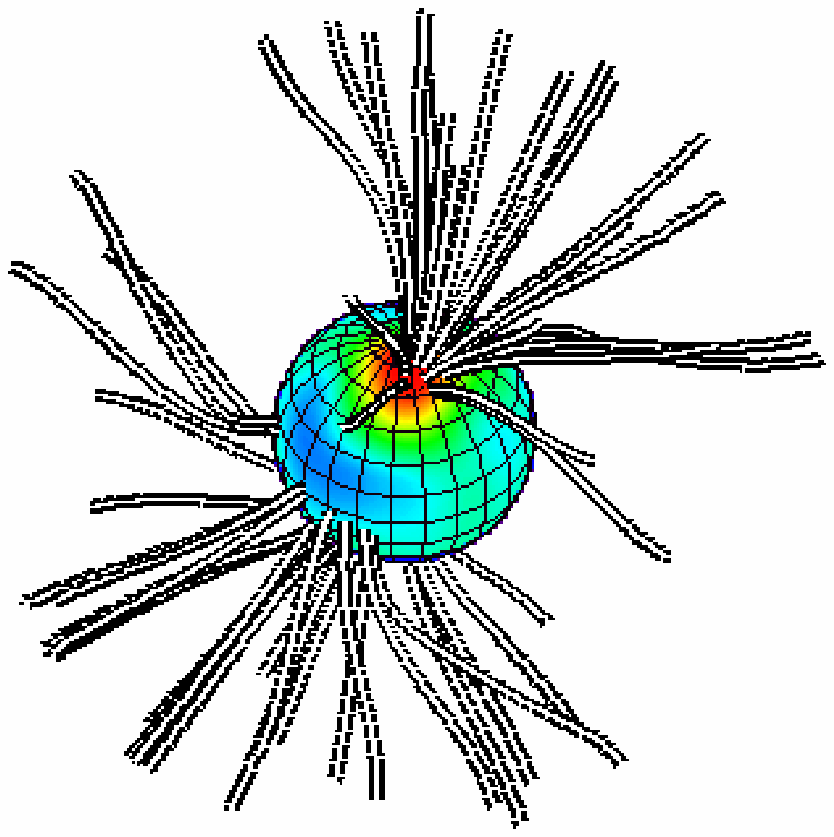,width=45mm}
                        } \\
        \end{tabular}
        \caption[]{Field extrapolations from observationally derived surface magnetograms of BP~Tau (top row) and
                  V2129~Oph (bottom row).  The left-hand panel shows the complex surface field, the middle panel the  
                  more well-ordered larger scale field, and the right-hand panel the open field lines.  The stellar surface
                  is coloured to show the polarity of the radial field component with red (blue) depicting positive (negative)
                  field regions.  The images are not to scale.}
        \label{field_extrapolations}
\end{figure*}

\subsection{Magnetic field measurements}
From Zeeman-Doppler imaging of V2129~Oph, \citet{don07} detected clear circular polarisation
signals in both photospheric absorption, and accretion related emission, line profiles.  
The temporal variations of the Zeeman signatures was dominated by 
rotational modulation.  In deriving a surface magnetogram it is possible to determine
how the magnetic energy is distributed between the different field modes of a complex multi-polar
field.  In other words it is possible to determine how strong the dipole component is relative to 
the quadrupole component, or the octupole component, or the hexadecapole component etc.  In the case of 
V2129~Oph the field is dominated by a $1.2\,{\rm kG}$ octupole, tilted at $\sim$20\degr with respect to 
the stellar rotation axis. The dipole component was found to be weak, with a polar strength of only
$0.35\,{\rm kG}$ and tilted at $\sim$30\degr \citep{don07}.  The surface field in the visible hemisphere of V2129~Oph is dominated 
by a $2\,{\rm kG}$ positive radial field spot at high latitude, with the footpoints of the accretion
funnel rooted in this region, but differs significantly from a dipole \citep{don07}.  We note
that due to the limited resolution achievable by Zeeman-Doppler imaging studies, derived surface magnetograms
likely miss the smallest scale field at the stellar surface.  The integrated affects of many flares due to magnetic 
reconnection events within such small scale field regions, is likely responsible for the ``quiescent'' 
level of X-ray of emission detected from T Tauri stars \citep{get08a}.      

Several authors have reported the detection of magnetic fields on BP~Tau (\citealt*{joh99b}; \citealt{joh07}; 
\citealt{don08a}) with a strong circular polarisation signal commonly detected in the accretion related HeI 5876{\AA } 
emission line (\citealt{joh99a}; \citealt{val04}; \citealt{sym05}; \citealt{chu07}), as well as in other accretion related
emission lines (e.g. the CaII IRT and HeI 6678{\AA }; \citealt{don08a}).     
Despite the concerns of \citet{chu07}\footnote{Some authors have argued that variations in the longitudinal (line-of-sight) 
field component, derived from polarisation detections in the HeI 5876{\AA } line, were attributable to rotational modulation
(e.g. \citealt*{val03}; \citealt{val04}).  
\citet{chu07} refutes such suggestions and argues that the field in the HeI 5876{\AA } line formation region is constantly 
evolving and restructuring on a timescale of only a few hours.  The ESPaDOnS/NARVAL spectropolarimetric data presented by 
\citet{don08a}, however, clearly show that although the HeI 5876{\AA } line is subject to intrinsic variability, its temporal 
evolution is dominated by rotational modulation.  This suggests the magnetic field in the HeI line formation region
remains stable on timescales of longer than a rotation cycle.}, \citet{don08a} have presented clear evidence that the 
Zeeman signatures in both photospheric and accretion related lines are dominated by rotational modulation, having 
monitored BP~Tau for a complete rotation cycle at two different epochs.  Interesting the large scale structure of 
BP~Tau's field was found to be similar despite the observing runs being separated by almost 11 months.  However, the 
magnetic features appear to have undergone an apparent phase shift of a quarter of a rotation phase, which was most
likely caused by a small error in the adopted stellar rotation period \citep{don08a}.\footnote{Due to the similarity of the derived 
surface magnetograms we only use the one derived from the \citet{don08a} February 2006 data
in this paper.}  Although BP~Tau's magnetic field is found to be complex, it is simpler than that of V2129~Oph.
The field of BP~Tau consists of both a strong dipole component, which at $1.2\,{\rm kG}$ is four times as strong
as that of V2129~Oph, and a strong octupole component of $1.6\,{\rm kG}$.  Both the octupole and dipole
moments are tilted by $\sim $10\degr with respect to the stellar rotation axis, but in different planes \citep{don08a}.   


\section{Field extrapolation}
A variety of field extrapolation techniques have been developed which allow the 
extrapolation of coronal magnetic fields from observationally derived surface magnetograms.
From the derived maps of the photospheric fields
of V2129~Oph and BP~Tau, we extrapolate their three-dimensional coronal field topologies using the 
potential field source surface model.  A potential field is one which is current-free, and
has the advantages over a full MHD model of faster computation speed and simplicity, and does not
require an assumption about an equation of state.  The disadvantages
of such a model, however, are that time-dependent and non-potential effects cannot be reproduced, and 
these may be important when considering the interaction between the stellar field and the disc.    
\citet{ril06} provide a comparison of the two techniques, as applied to the Sun, and find that 
often the potential field model produces results that closely match MHD models, with \citet{liu07} 
coming to the same conclusion for the case of stable solar active regions.  For cTTSs however, the main source
of non-potentiality in the field is likely to be caused by the interaction between the stellar magnetosphere
and the disc, which cannot be modelled with the static field structures considered here.

Alternative field extrapolation techniques from Zeeman-Doppler images have been considered by 
\citet{hus02}, who compared the difference between a non-potential current carrying magnetic field
model, and the potential field model.  By considering field extrapolations from a magnetogram of the 
young rapid-rotator AB~Dor,  they found that the ratio of free energy in the potential to that
in the non-potential model differed by up to 20\% close to the stellar surface.  However, this difference 
decayed rapidly with height above the star, with only 1\% difference at $1R_{\ast}$ above the photosphere, 
indicating little difference in the field structures derived from the potential and non-potential field models.  
The potential field source surface model should therefore be adequate to allow us to address the question of
whether the surface magnetograms of V2129~Oph and BP~Tau are consistent with the observed locations 
of accretion funnels and accretion hotspots, as well as to compare the large scale field topologies with
that of a simple dipole.

As we have assumed that the field is potential then $\nabla \times \bmath{B} = 0$.  This condition can be satisfied 
by writing the field in terms of a scalar flux function $\Psi$
\begin{equation}  
\bmath{B} = -\nabla \Psi.
\label{b}
\end{equation}
The field must also satisfy Maxwell's equation, $\nabla \cdot \bmath{B}=0$, and therefore 
$\Psi$ must satisfy Laplace's equation, $\nabla^2 \Psi = 0$.  This has 
a standard solution of the form,
\begin{equation}
\Psi= \sum_{l=1}^{N} \sum_{m=-l}^{l} \left
[a_{lm}r^l+b_{lm}r^{-(l+1)} \right ] P_{lm}(\theta) {\rm e}^{{\rm i}m\phi},
\label{flux_function}
\end{equation}
where $P_{lm}$ denote the associated Legendre functions and the coefficients $a_{lm}$ and $b_{lm}$
are determined from the boundary conditions.  The first boundary condition is that the strength
of the radial component of the field at the stellar surface is that which is derived from 
the Zeeman-Doppler maps of the star.  The second condition is that at some height above the stellar
surface $R_S$, known as the {\it source surface}, the field becomes radial and hence $B_{\theta}(R_S)=B_{\phi}(R_S)=0$, 
emulating the effect of the corona blowing open field lines to form a stellar wind \citep{alt69}.  
Thus, from (\ref{b}) it is possible to determine the magnetic field components, $B_r$, 
$B_{\theta}$, and $B_{\phi}$, and therefore a field vector, at any point within the 
stars corona (see \citet{gre06a} for the full expressions).  In order to extrapolate the 
field we used a modified version of a code originally developed by \citet*{van98}, and applied
specifically to T Tauri stars by \citet{jar06} and \citet{gre06a,gre06b,gre07}.  

The field extrapolations showing the smaller scale, larger scale, and open field structure of
V2129~Oph and BP~Tau, are shown in Fig. \ref{field_extrapolations}.  For both stars we have set the source surface to be at the 
corotation radius (see \S2.1), which at this stage serves only to illustrate the difference between the 
complex and loopy surface field regions, and the well-ordered larger scale field.  Later we discuss
how variations in the source surface radius affects the structure of the field.  It is clear from 
Fig. \ref{field_extrapolations} that for both stars the largest
scale field resembles a slightly tilted dipole, especially in the case BP~Tau.  But how does the larger scale
field extrapolated from observationally derived surface magnetograms, compare to a simple 
dipole magnetic field?   


\subsection{Comparison with a large scale dipole field}
The complex field structures obtained by field extrapolation of both V2129~Oph, and in particular
BP~Tau, appear to indicate that the larger scale field is well ordered, and much simpler than the 
loopy and compact surface field regions.  In this section we compare the larger scale field
topology with that of a modified dipolar field.   Taking the $l=1$, $m=0$ component of equation 
(\ref{flux_function}) and imposing the boundary conditions
\begin{eqnarray}
B_r(R_{\ast}) &=& \frac{2\mu}{R_{\ast}^3}\cos{\theta} \\
B_{\theta}(R_S) &=&  B_{\phi}(R_S) = 0,
\end{eqnarray}
allows the construction of a dipole field with a source surface,
where $\mu = R_{\ast}^3 B_{\ast, pole}/2$ is the dipole moment and $B_{\ast, pole}$
the polar strength of the dipole.  The inclusion of a source surface introduces 
a modification to the standard dipole field components,
\begin{eqnarray}
B_r      &=& \frac{2\mu\cos{\theta}}{r^3}\left( \frac{r^3+2R_S^3}{R_{\ast}^3+2R_S^3}\right)\\
B_\theta &=& \frac{\mu\sin{\theta}}{r^3}\left( \frac{-2r^3+2R_S^3}{R_{\ast}^3+2R_S^3}\right)\\ 
B_\phi   &=& 0,
\end{eqnarray}
which are recovered in the limit of $R_S \to \infty$.  Individual field lines 
can be described by,
\begin{equation}
\sin^2{\theta} = \frac{\xi r}{r^3 + 2R_S^3},
\end{equation}
where $\xi$ is a constant along a particular field line, such that different values of $\xi$ correspond to 
different field lines.  The last closed field line passes through the stars equatorial plane ($\theta = \pi/2$)
at $r=r_m=R_S$, and therefore $\xi=3R_S^2$.  Such a field line connects to the stellar surface at a co-latitude of 
$\theta_m$ where
\begin{equation}
\theta_m = \sin^{-1}\left( \sqrt{\frac{3R_S^2R_{\ast}}{R_{\ast}^3 + 2R_S^3}}\right).
\label{theta_m}
\end{equation}     
The structure of such a modified dipolar field is illustrated in fig. 1 of \citet{jar06}.  We note that
the dipole components of the fields of both BP~Tau and V2129~Oph are slightly tilted with respect to the 
stellar rotation axis (10\degr for BP~Tau and 30\degr for V2129~Oph).  We account for this small tilt 
below when comparing the complex field structures to that of the dipolar fields.  \citet{mah98} provide
analytic expressions for calculating the components of an arbitrarily tilted dipole, which we adapt in 
order to include the source surface, and calculate the tilted dipole field components numerically using
our field extrapolation code.  

\begin{figure*}
        \def\subfigtopskip{4pt}
        \def\subfigbottomskip{4pt}
        \def\subfigcapskip{2pt}
        \centering
        \begin{tabular}{cc}
                \subfigure{
                        \label{lw_bptau_small}
                        \psfig{figure=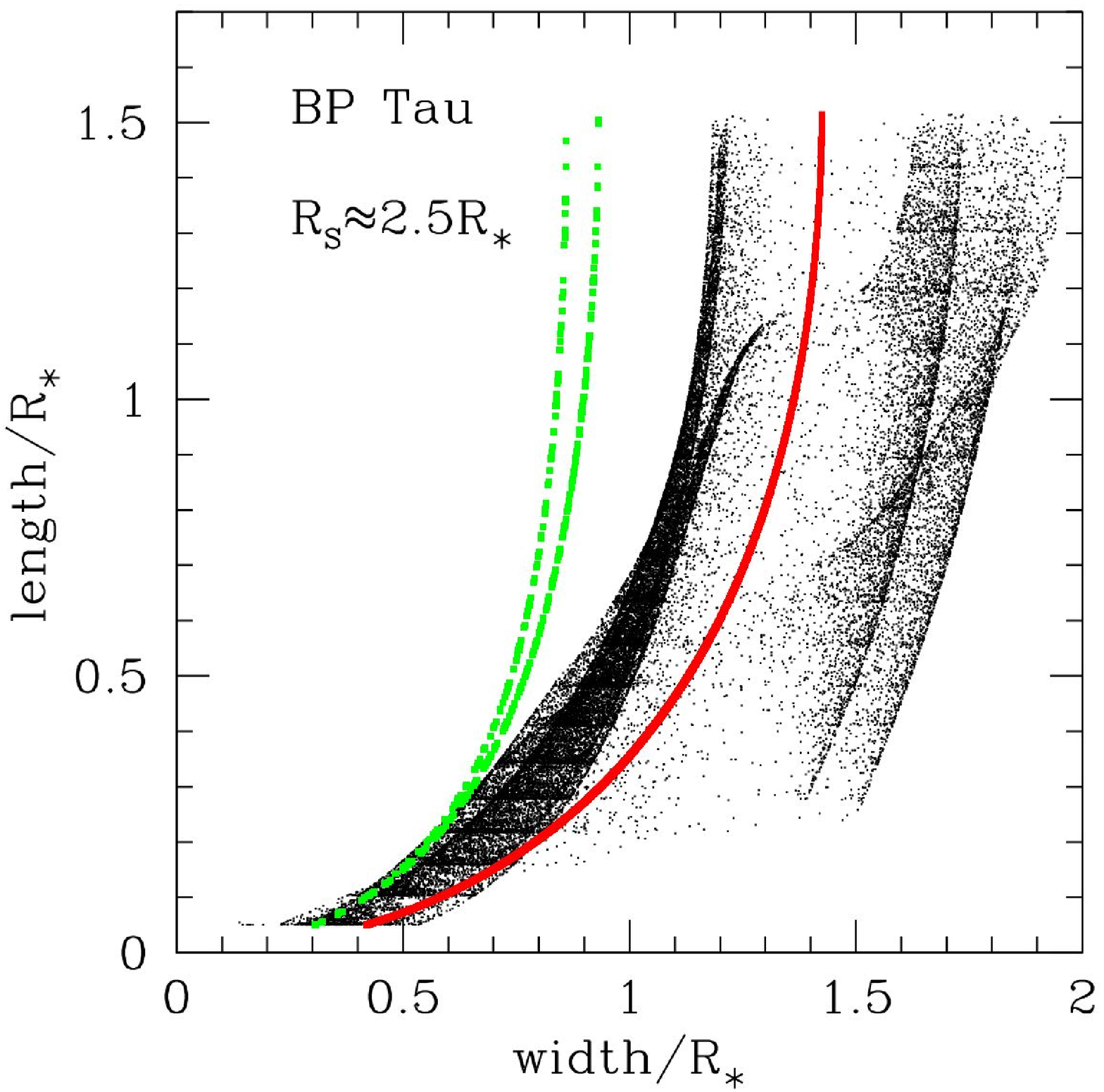,width=80mm}
                        } &
                \subfigure{
                        \label{lw_v2129_small}
                        \psfig{figure=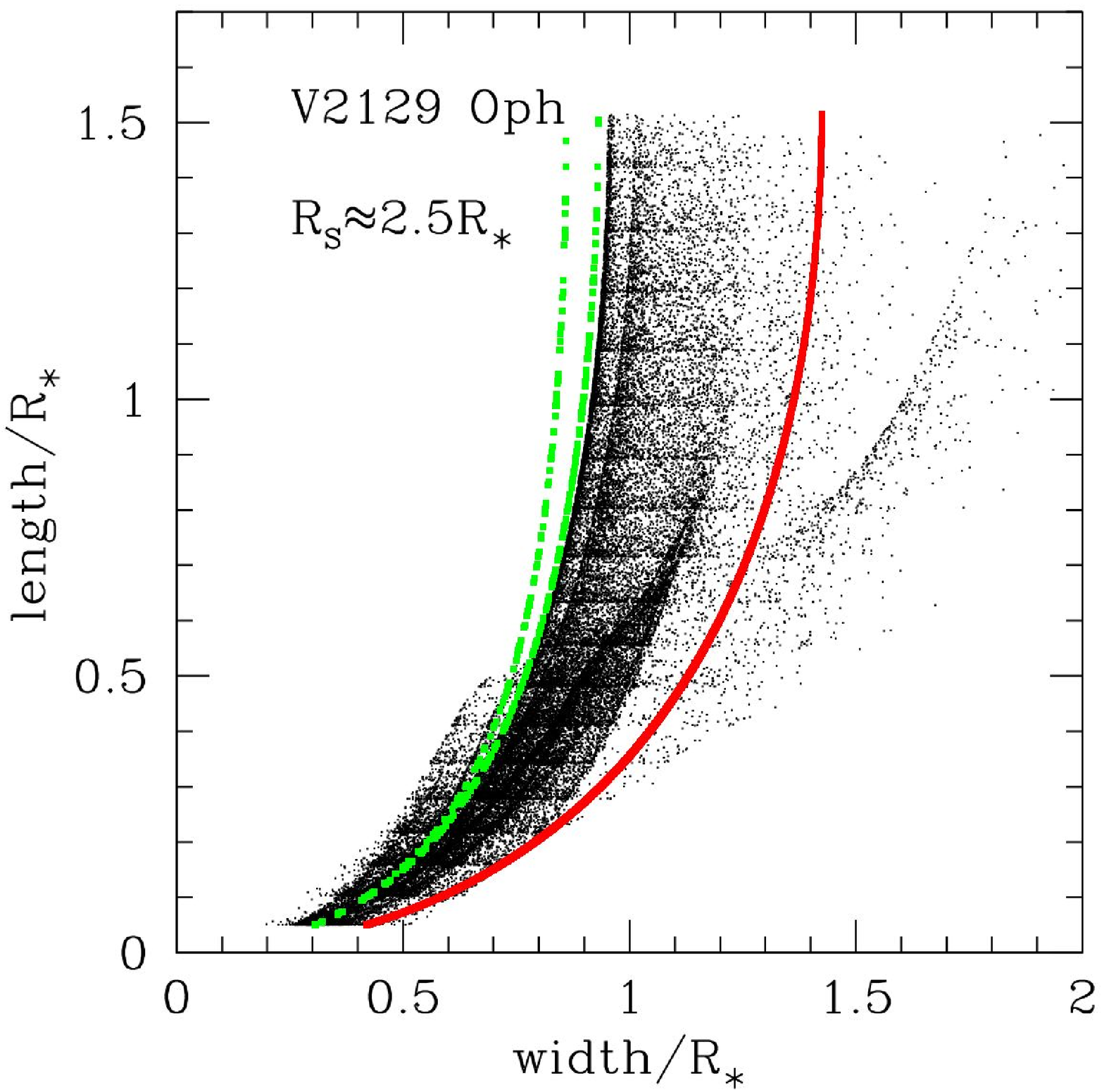,width=80mm}
                        }  \\
        \end{tabular}
        \caption[]{Plots of field line length (maximum radial distance above the stellar surface) against field line width (the distance along a 
                  segment of the great circle connecting the field line footpoints on the stellar surface) for BP~Tau (left panel) and V2129 Oph (right panel). 
                  A source surface radius of $2.5\,{R_{\ast}}$ has been assumed.  The black points are for the extrapolated fields,
                  the red points indicate the behaviour of a dipole field, and the green points an octupole field.}
        \label{lw_small}
\end{figure*} 

\begin{figure*}
        \def\subfigtopskip{4pt}
        \def\subfigbottomskip{4pt}
        \def\subfigcapskip{2pt}
        \centering
        \begin{tabular}{cc}
                \subfigure{
                        \label{lw_bptau_big}
                        \psfig{figure=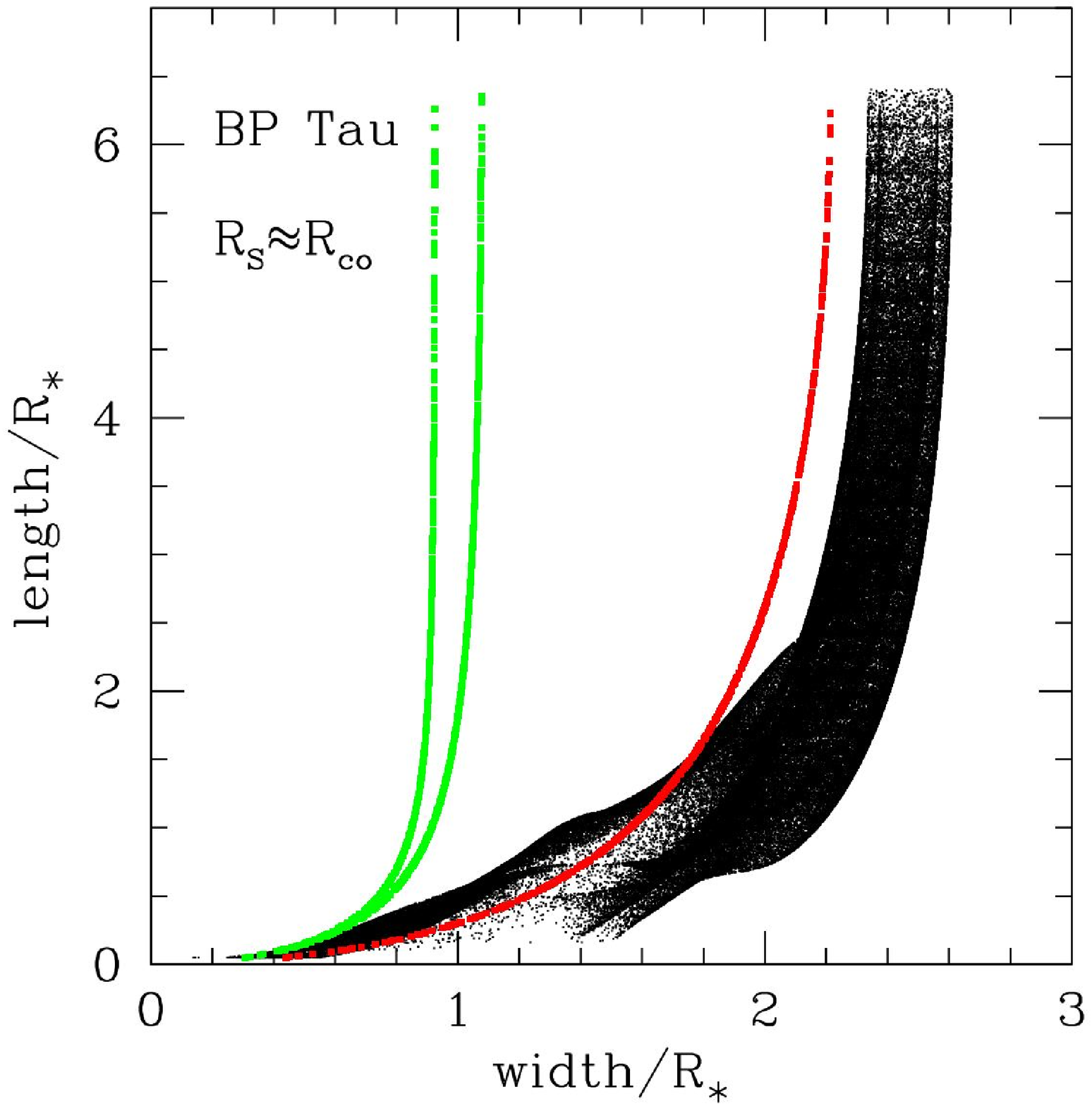,width=80mm}
                        } &
                \subfigure{
                        \label{lw_v2129_big}
                        \psfig{figure=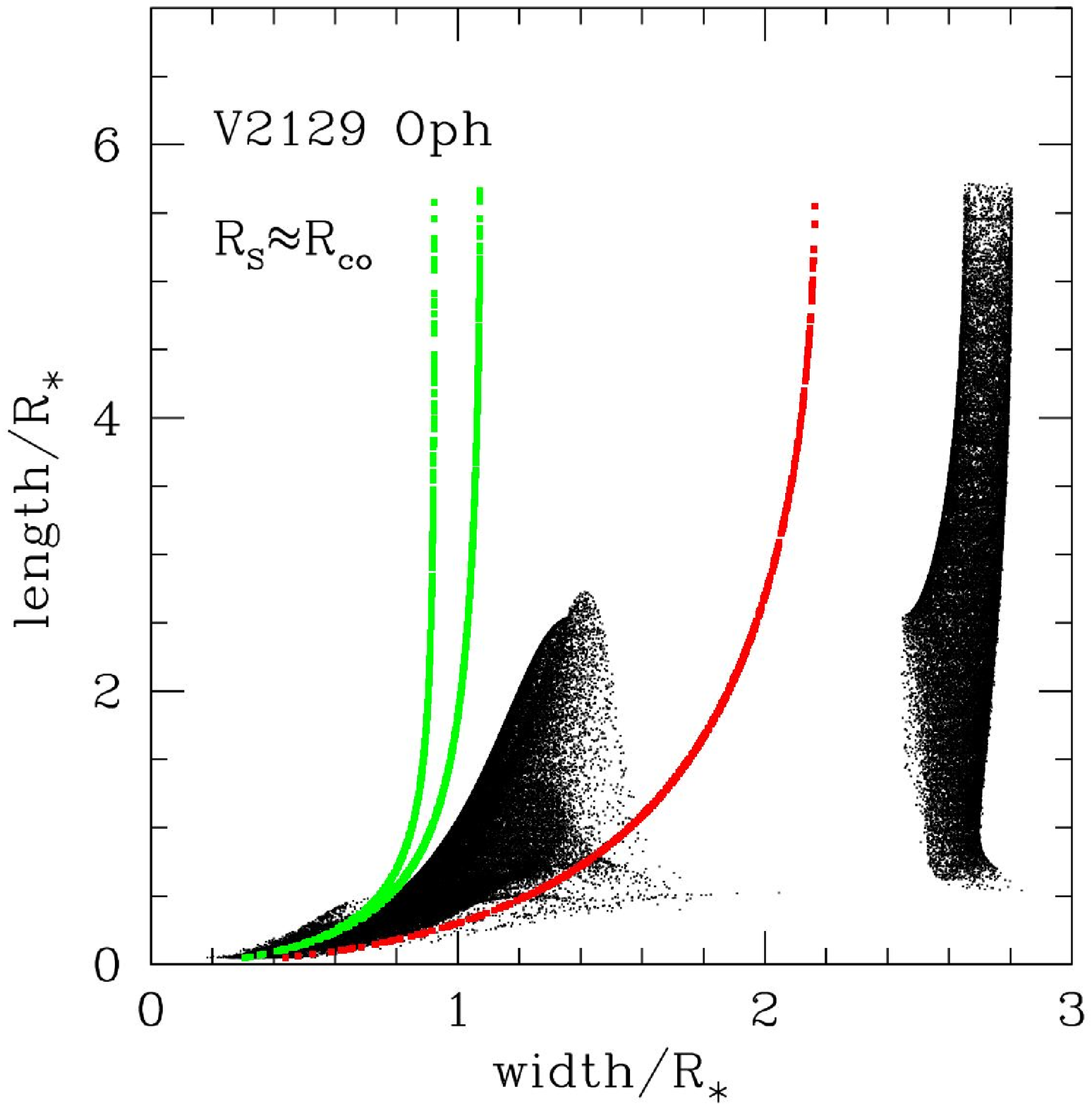,width=80mm}
                        }  \\
        \end{tabular}
        \caption[]{The same as Fig. \ref{lw_small}, but assuming that the source surface is at approximately the corotation radius,
                  which is $7.4\,R_{\ast}$ for BP~Tau (left panel) and $6.7\,R_{\ast}$ for V2129~Oph (right panel).}
        \label{lw_big}
\end{figure*} 

The black points in Figs. \ref{lw_small} and \ref{lw_big} show field line length against width calculated
from the field extrapolations of the magnetic fields of BP~Tau and V2129~Oph, for source
surface radii of $R_S = 2.5\,{R_{\ast}}$ and $R_S \approx R_{co}$.  At this stage such plots
are used only to illustrate by how much the field structure departs from a simple dipole (red points
in Figs. \ref{lw_small} and \ref{lw_big}).  We define the length of a field line as being the radial 
distance measured from the stellar surface 
to the maximum height of the field line.  The width of a field line is defined as the distance
along the segment of the great circle connecting the field line footpoints on the stellar surface.  For a field line 
with footpoints at {\it co-latitudes} and longitudes $(\theta_1,\phi_1)$ and $(\theta_2,\phi_2)$
the width, $w$, can be calculated using the Haversine formula,
\begin{eqnarray}
h &=& \sin^2 \left( \frac{\Delta \theta}{2}\right) + \sin \theta_1 \sin \theta_2 \sin^2 \left( \frac{\Delta \phi}{2}\right)\\
w &=& 2\sin^{-1} \left( \sqrt{h}\right),
\end{eqnarray}
where $\Delta \phi = \phi_2 - \phi_1$ is the longitudinal separation between the footpoints,
$\Delta \theta = |\theta_2 - \theta_1|$ is the difference in co-latitudes, and where all angles are
measured in radians and all distances in units of stellar radii.    

By considering progressively longer field line loops, Figs. \ref{lw_small} and \ref{lw_big} show that for a dipole field the 
field line width eventually tends asymptotically to a certain value which depends on the 
choice of source surface.  This is as expected, as by equation (\ref{theta_m}) the co-latitude of the footpoint of
the longest dipole field line depends on the choice of source surface $R_S$, with $\theta_m$ being smaller for a
larger $R_S$ (and for an aligned dipole field line the latitudinal footpoint separation is 
$\Delta \theta = \pi - 2\theta_m$ radians).  We also show in Figs. \ref{lw_small} and \ref{lw_big} the behaviour of 
an octupole magnetic field (green points).  For an axial octupole ($l=3, m=0$) the width of the closed field line
loops about the stellar equator is less than those at higher latitudes (see the discussion in \citealt{wil87}), which
gives rise to two separate lines in the length vs width plots.   

For a small source surface (see Fig. \ref{lw_small}) the longer field lines on BP~Tau are
typically wider than those on V2129~Oph.  This reflects the strength of the dipole components in the two
stars, which is about 4-times stronger on BP~Tau than V2129~Oph (stars with a strong dipolar field component would 
typically have wider field lines, connecting the different polarity poles in opposite hemispheres), and 
also the dominantly octupolar nature of V2129~Oph's field.  In both cases, however, the fields are more complex than 
a dipole.  For the case of a large source surface (see Fig. \ref{lw_big}) the field structure of BP~Tau follows
a similar trend to the dipole field, as do the longest field lines on V2129~Oph.  
However, many of the smaller scale field lines on V2129~Oph are not as wide as the dipolar field
lines.  This suggests that the smaller scale field on V2129~Oph is complex and multi-polar with the field strength 
decaying more rapidly with height than on BP~Tau; we return to this point in \S5.
For V2129~Oph the smaller scale field (see Fig. \ref{lw_small}) is more closely matched to an octupole field.
Also, some of the shortest width field lines are longer than those of an octupole, indicating a contribution
to the surface field from the higher order field modes.  Indeed, for V2129~Oph there is also a significant amount
of magnetic energy in the $l=5$ field mode \citep{don07}.   

The field structure of V2129~Oph is particularly striking when a large source surface is considered (Fig. {\ref{lw_big}, right panel).
The surface magnetic field of V2129~Oph is dominated by large 2~kG positive radial field spot at high latitude 
in the visible hemisphere, and likely a similar negative field region in the unseen hemisphere \citep{don07}.  
While the surface field region consists of many small width loops, the larger scale field comprises 
of the longer magnetic loops which connect to the high latitude field regions.
This behaviour is not apparent when considering V2129~Oph with a smaller source surface, as 
the co-latitude at which the last closed field line connects to the stellar surface is larger, by
equation (\ref{theta_m}), which means that field lines cannot connect to the large high latitude radial field spots. 
With a large source surface the longest field lines are always wider than those of a dipole, 
suggesting that the structure of the field lines are being affected by the strong field regions 
close to the stellar surface; with such 
field lines connecting at smaller co-latitudes than a dipole field.  This effect is more apparent for V2129~Oph,
where the larger scale field lines are ``squeezed'' by the strong dominantly octupolar field close to the stellar
surface.  This forces such field lines to have larger widths at a given length.  This is also the case for the field
of BP~Tau (see Fig \ref{lw_big}, left panel), however, the effect is less due to the strength of octupole component
relative to the dipole component being over 2.5 times less for BP~Tau than for V2129~Oph.  The structure of the largest
scale field lines therefore depart from the path followed by purely dipolar field lines.  It is such larger scale 
loops which are likely 
to be carrying material from the disc to the star.  Therefore the location of 
accretion hotspots on the stellar surface will differ from what is expected from dipolar accretion, even though
the larger scale field is relatively well-ordered and similar to a dipole.   
In the following sections we look more closely at the structure of the open and accreting field.  


\section{The Open Field}

\begin{figure}
  \centering
  \psfig{file=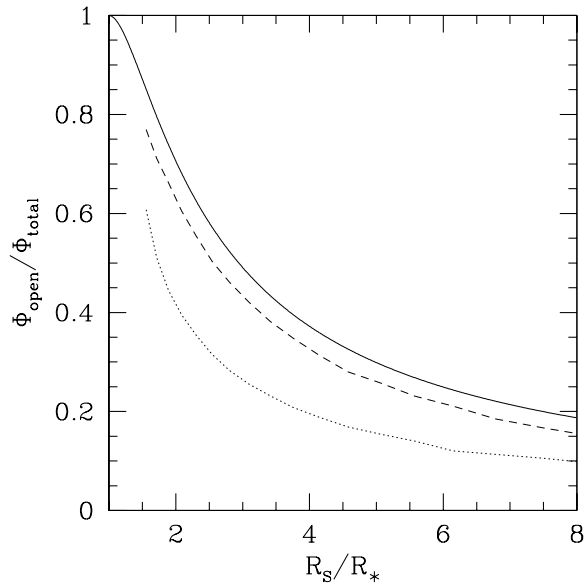,width=80mm}
  \caption{The variation of the ratio of open to total flux through the stellar surface for a
          range of source surface radii.  The solid line represents the modified dipole magnetic
          field, the dashed line the field of BP~Tau and the dotted line the field of V2129~Oph.  There is
          less fractional open flux when considering magnetic fields with a realistic degree of complexity.}
  \label{fluxratio}
\end{figure}

\begin{figure*}
        \def\subfigtopskip{4pt}
        \def\subfigbottomskip{4pt}
        \def\subfigcapskip{2pt}
        \centering
        \begin{tabular}{cc}
                \subfigure{
                        \label{fluxopen}
                        \psfig{figure=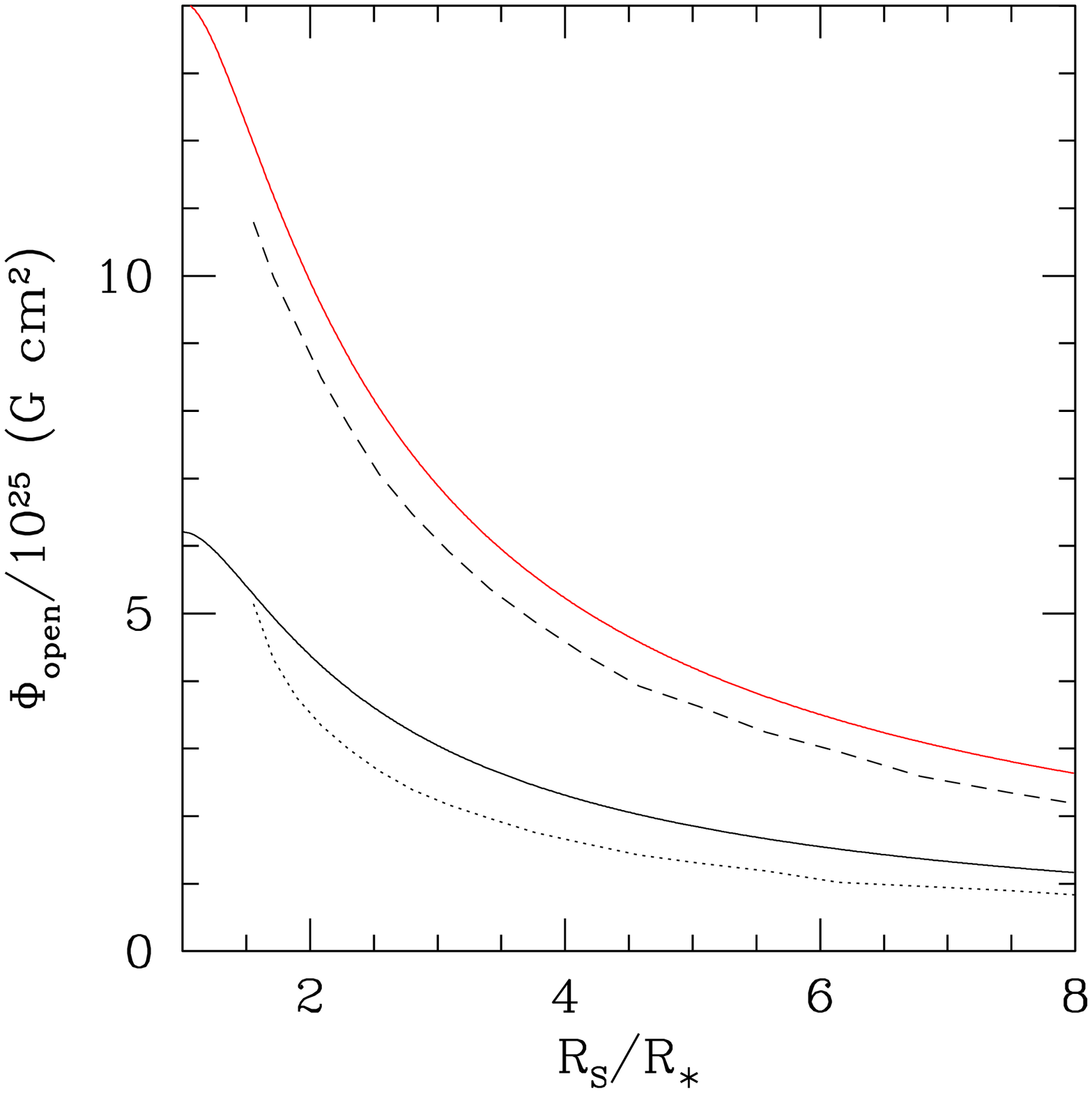,width=80mm}
                        } &
                \subfigure{
                        \label{fluxclosed}
                        \psfig{figure=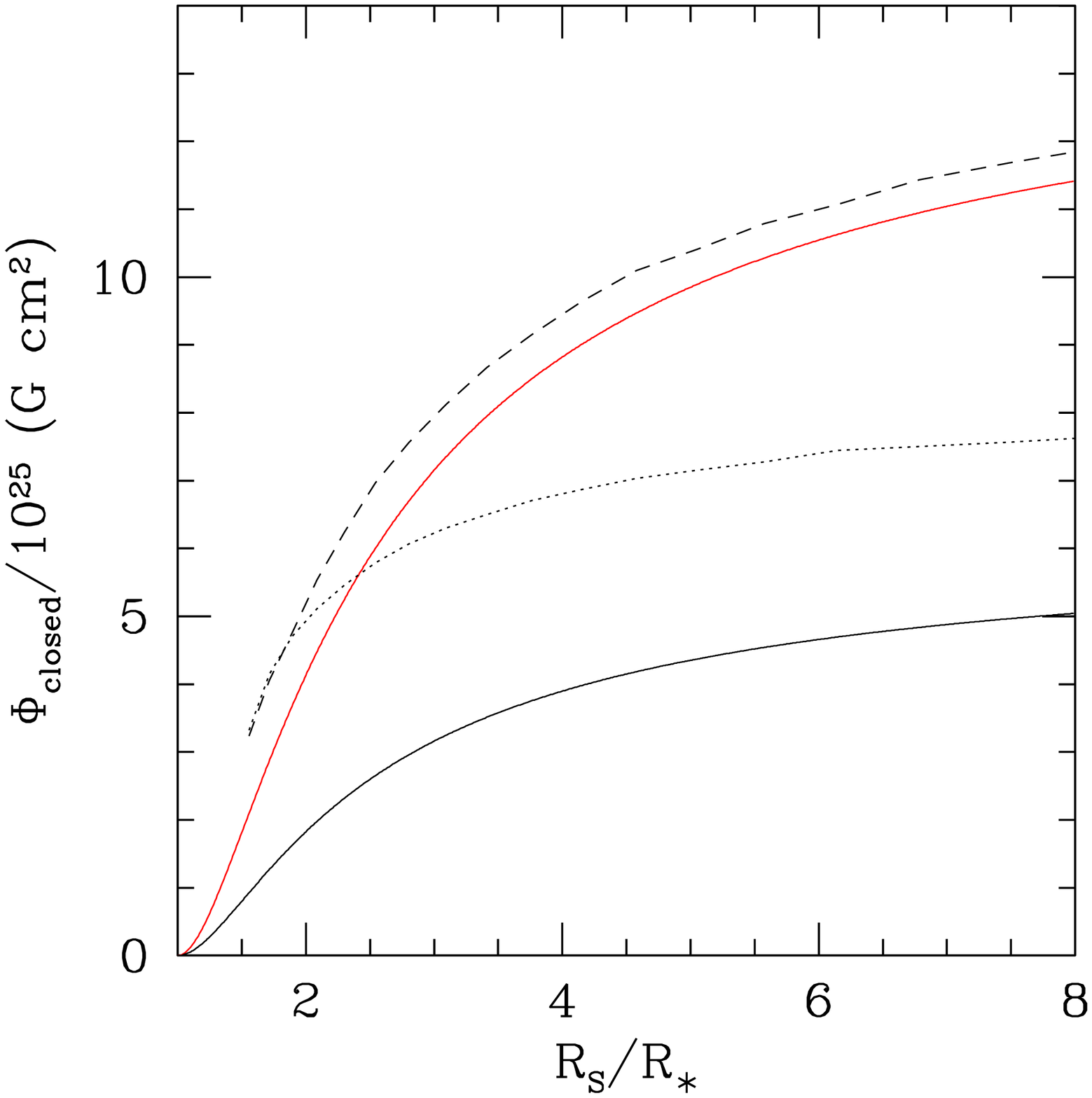,width=80mm}
                        }  \\
        \end{tabular}
        \caption[]{The variation of open and closed flux (left and right panels respectively)
                   through the stellar surface for a range of source surface radii.  The solid lines represent modified
                   dipole magnetic fields, with the red (black) line representing a polar field
                   strength of 1.2 (0.35)$\,{\rm kG}$.  The dashed line represents the 
                   field of BP~Tau and the dotted line the field of V2129~Oph.  There is less open flux, but more 
                   closed flux, when considering the complex magnetic fields.}
        \label{fluxes}
\end{figure*}

The amount of open flux passing through the stellar surface is important for models of stellar
winds, which for cTTSs may remove the angular angular momentum which would be transfered to the star due
to accretion \citep{mat05b}.  It is therefore of
interest to compare the amount of open flux for V2129~Oph and BP~Tau with that of the modified
dipole magnetic field.  Since the magnetic torque in a stellar wind does not depend on the polarity of the 
field, we consider here the absolute value of the flux.  The open flux through the stellar surface is given by
\begin{equation}
\Phi_{open} = R_{\ast}^2 \int\int |B_{r,open}(\theta,\phi)| d\Omega.
\label{openflux}
\end{equation}
Using equation (4) an expression for $\Phi_{open}$ can be derived for our modified dipole field,
\begin{eqnarray}
\Phi_{open} &=& 2R_{\ast}^2 \int_0^{2\pi}d\phi \int_0^{\theta_m} \frac{2\mu}{R_{\ast}^3} \cos{\theta}\sin{\theta}d\theta \\
            &=& 2\pi R_{\ast}^2 B_{\ast, pole} \sin^2{\theta_m}.
\label{openfluxdipole}
\end{eqnarray}
where the additional factor of 2 accounts for the open field in each hemisphere.  A similar
expression can be derived for the closed flux through the stellar surface,
\begin{eqnarray}
\Phi_{closed} &=& 2R_{\ast}^2 \int_0^{2\pi}d\phi \int_{\theta_m}^{\pi/2} \frac{2\mu}{R_{\ast}^3} \cos{\theta}\sin{\theta}d\theta \\
            &=& 2\pi R_{\ast}^2 B_{\ast, pole} \cos^2{\theta_m}.
\label{closedfluxdipole}
\end{eqnarray}
Thus for the modified dipole field the ratio of open flux to the total flux through the 
stellar surface is,
\begin{equation}
\frac{\Phi_{open}}{\Phi_{total}} = \sin^2{\theta_m} = \frac{3R_S^2R_{\ast}}{R_{\ast}^3 + 2R_S^3},
\label{fluxratioeq}
\end{equation}
where the total flux $\Phi_{total} = \Phi_{open} + \Phi_{closed}$.  The solid line in Fig. \ref{fluxratio}
shows the variation of the fractional open flux $\Phi_{open}/\Phi_{total}$ for a range of different source surface radii 
$R_S$ assuming a modified dipolar magnetic field.  For large source surface radii the colatitude of the footpoint of
the last closed field line is close to the pole of the star, and therefore the open flux through the stellar
surface is less.  As the source surface radius is decreased more of the closed field is converted to open
field along which a (stellar) wind may be launched.  

For the magnetic fields extrapolated from the surface magnetograms we divide the stellar surface 
into a series of small grid cells.  Within each grid cell we take the average of the magnitude of $B_r$,
at the footpoint of the open field lines and multiply this by the area of the cell in order
to calculate the open flux from that individual cell.  If a cell contains the footpoints of both
open and closed field lines we weight the area of that cell accordingly.  For example, a cell containing
50$\%$ open field, is assumed to contribute 50$\%$ of it's area to the open flux.  The total open flux through the 
stellar surface is then obtained by summing over all the cells, while the total closed flux is calculated in a 
similar way.  We find that there is less fractional open flux through the surface of V2129~Oph and BP~Tau 
(dotted and dashed line in Fig. \ref{fluxratio}) compared with the dipole field.
This is consistent with our argument in the previous section that the structure of the 
larger scale field (and in particular that of V2129~Oph) is influenced by the strong field regions closer to the star.  
As the larger scale field lines are wider than those of a dipole, their footpoints are closer to the 
pole (see Fig. \ref{lw_big}).  Thus there is less area of the stellar surface available from which
a stellar wind can be launched along the open field.    

The lower fractional open flux for the complex fields, can be understood by considering 
how the open and closed flux changes as the source surface location is varied.  This is illustrated in 
Fig. \ref{fluxes}, where the solid lines represent dipole magnetic fields.  In order to compare the fluxes of the complex
fields with a dipole, we have assumed that the dipole has a polar strength equal to that 
of the measured dipole components of V2129~Oph and BP~Tau (0.35$\,{\rm kG}$ and 1.2$\,{\rm kG}$ respectively).  Both stars
have slightly less open flux compared to a dipole, but more closed flux.  There is significantly more closed flux
for V2129~Oph compared to a dipole than for BP~Tau.  This is simply a reflection of the more complex nature of V2129~Oph's 
magnetic field, and is the largest factor in explaining why the fractional open flux for this star is well below the 
dipole case (see Fig. \ref{fluxratio}). 

The main result of this section is that there is less fractional open flux for the complex field geometries, compared to
a pure dipole case.  This is result of the complex fields having both lower open flux, and, in particular, higher closed flux. 
This is understandable since a more complex field geometry is characterised by more numerous and smaller closed 
magnetic loops.  


\section{The accreting field}
It has long been suspected that the magnetic fields of cTTSs are 
more complex than a simple dipole, with, for example, \citet{edw94} arguing that 
the common detection of inverse P-Cygni profiles was evidence for 
multiple accreting loops in a field configuration more complex than a dipole.  
Until very recently, however, only indirect clues were available to the complex nature of 
T Tauri magnetic fields.  The detection of rotationally modulated X-ray emission in a small,
but significant, sample of T Tauri stars indicated that X-ray emitting plasma was confined
to compact field regions in-homogeneously distributed about the stellar surface \citep{fla05}.  
Such modulated X-ray emission would not occur if the surface field was dipolar.  Further 
clues to the complex nature of T Tauri surface fields comes from the common failure to 
detect polarisation signals in photospheric absorption lines, despite strong average fields
being measured from the Zeeman broadened, and unpolarised,  intensity spectra \citep{val04}.
This suggested that the surface of T Tauri stars were covered in many regions of opposite polarity.
Spectral lines forming in such magnetic regions would be polarised in the opposite sense, 
leading to a cancellation of the polarisation signal along the line-of-sight to the observer.
However, often a strong polarisation signal is detected in accretion related emission lines
(e.g. \citealt{joh99a}) which is modulated due to the stellar rotation \citep{val04}.  This 
suggests that such emission lines are forming in regions of the stellar atmosphere which are permeated by 
field lines of a single magnetic polarity 
which remains stable as the star rotates.  In other words, despite the clues to the complex nature of 
T Tauri surface fields, the accreting field lines are likely anchored in single polarity regions, and the field 
connecting to such regions may be well-ordered and more like a dipole.   

The resolution achievable with ESPaDOnS (and its twin instrument NARVAL) has allowed the detection
of polarisation signals in both photospheric absorption, and accretion related emission, lines, and has
confirmed the complex nature of magnetic fields of cTTSs \citep{don07,don08a}.  However, magnetospheric
accretion models developed thus-far have yet to account for the true complexity of the stellar
magnetic field.  The first steps in considering how realistic multi-polar fields would 
influence accretion from the inner disc were taken by \citet{gre05,gre06a} and \citet{jar06}.  More
recently the 3D MHD simulations of \citet{lon07,lon08} have considered a quadrupole-dipole composite
field.  However, simulations incorporating generalised multi-polar magnetic fields are 
now required, and will shed new-light on the star-disc interaction. 

For the stars where magnetic surface maps have been obtained to date the bulk of accretion
in the visible hemisphere is channelled into a single magnetic spot at high latitude \citep{don07,don08a}.  
In order to explain the location of high latitude hotspots, the source surface has to be large enough
to allow field lines to connect between the high latitude radial field spots close to the poles.  The 
source surface radius represents the extent of the closed stellar magnetosphere and independently
obtained observations provide justification for setting the source surface to be at least equal 
to the corotation radius.  \citet{get08a,get08b} have analysed flares detected during the {\it Chandra} Orion 
Ultradeep Project, using a new technique that allows the modelling of flaring events which are much
fainter than those accessible by traditional parametric flare modelling techniques (e.g. \citealt{fav05}).  By 
calculating the lengths of the magnetic loops containing the flares, \citet{get08b} have found evidence for 
magnetic structures that extend well beyond the corotation radius radius in weak-line T Tauri stars, however 
for classical T Tauri stars the magnetic loops are all confined to within the corotation radius.  This
suggests the presence of a disc is limiting the extent of the closed stellar magnetosphere of accreting T Tauri 
stars - as previously predicted by \citet{jar06}.  The \citet{get08b} results therefore provide strong evidence 
that the stellar magnetosphere in classical T Tauri stars never extends to beyond the corotation radius, suggesting
that setting the source surface to be at corotation is a reasonable assumption.  However, the location of the disc 
truncation radius may still be closer to the stellar surface.  With a large source surface field lines are able
to connect to the high latitude field regions, even well within the corotation radius.  
Thus if the disc is truncated at, or even within, the corotation radius, then
accretion may proceed along field lines which connect
to a high enough latitude in order to explain the observed hotspot locations.
Therefore an interesting question to ask is how would
the disc truncation radius differ between dipole and more complex magnetic field models?


\subsection{The inner disc radius}
The question of where the disc is truncated, and therefore the structure of the magnetic field
threading the disc and carrying accreting gas, remains a major problem for accretion models.   It is still 
unknown if the disc is truncated in the vicinity of the corotation radius, the assumption of traditional 
accretion models (e.g. \citealt{kon91}), or whether it extends closer to the stellar surface
(e.g. \citealt{mat05b}).  As in \citet{gre07} we assume that accretion occurs over a range of radii within the 
corotation radius.  This is equivalent to the approach taken previously 
by \citet*{har94}, \citet*{muz01}, \citet{sym05}, \citet{aze06} and \citet*{kur06} who have 
demonstrated that such an assumption broadly reproduces observed spectral line profiles and variability.  
It should also be noted that the accreting field geometries which we consider here are only snap-shots in time, 
and in reality will evolve due to the interaction with the disc. 

\begin{figure*}
        \def\subfigtopskip{4pt}
        \def\subfigbottomskip{4pt}
        \def\subfigcapskip{2pt}
        \centering
        \begin{tabular}{cc}
                \subfigure{
                        \label{rt_bptau}
                        \psfig{figure=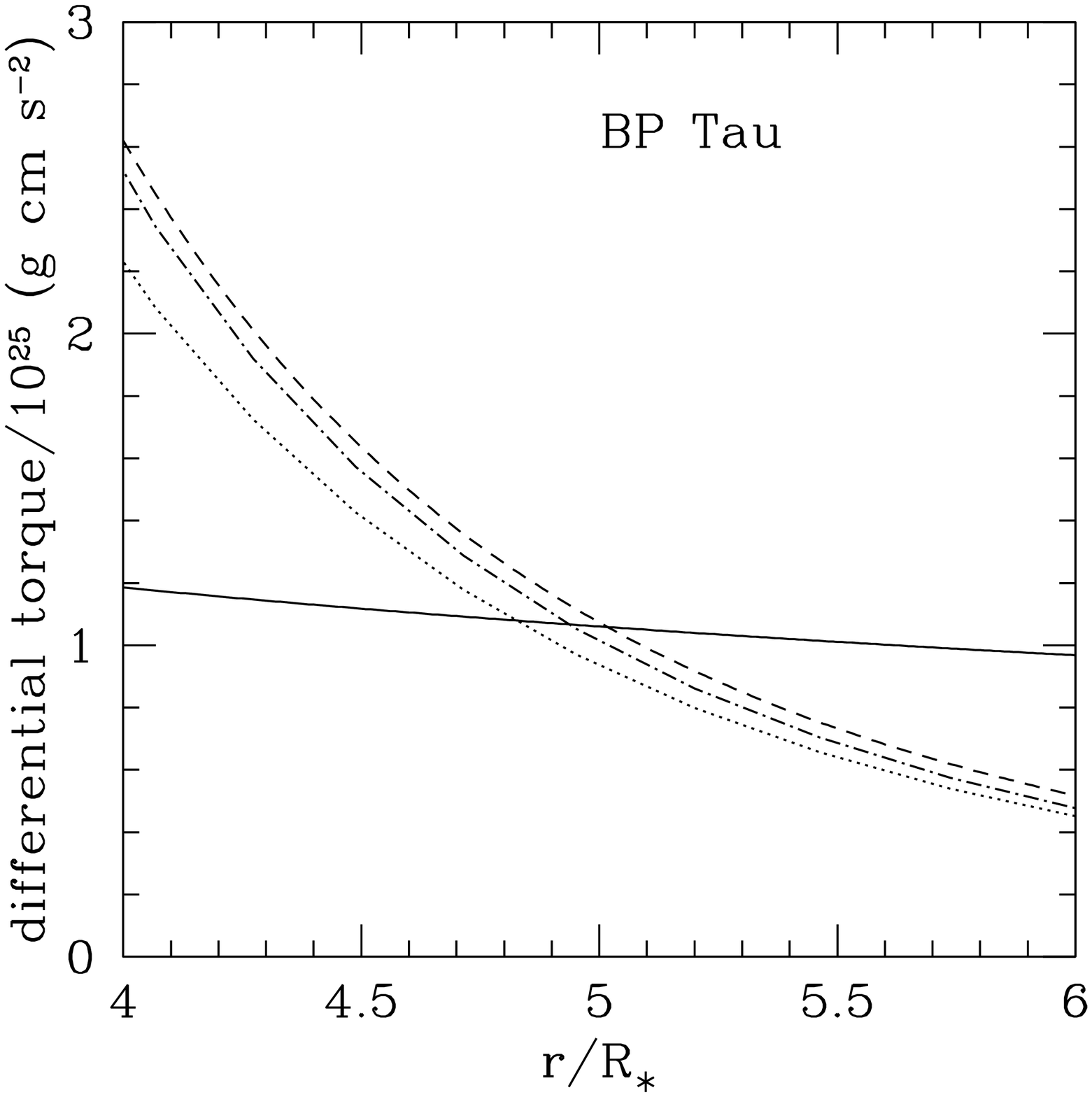,width=80mm}
                        } &
                \subfigure{
                        \label{rt_v2129oph}
                        \psfig{figure=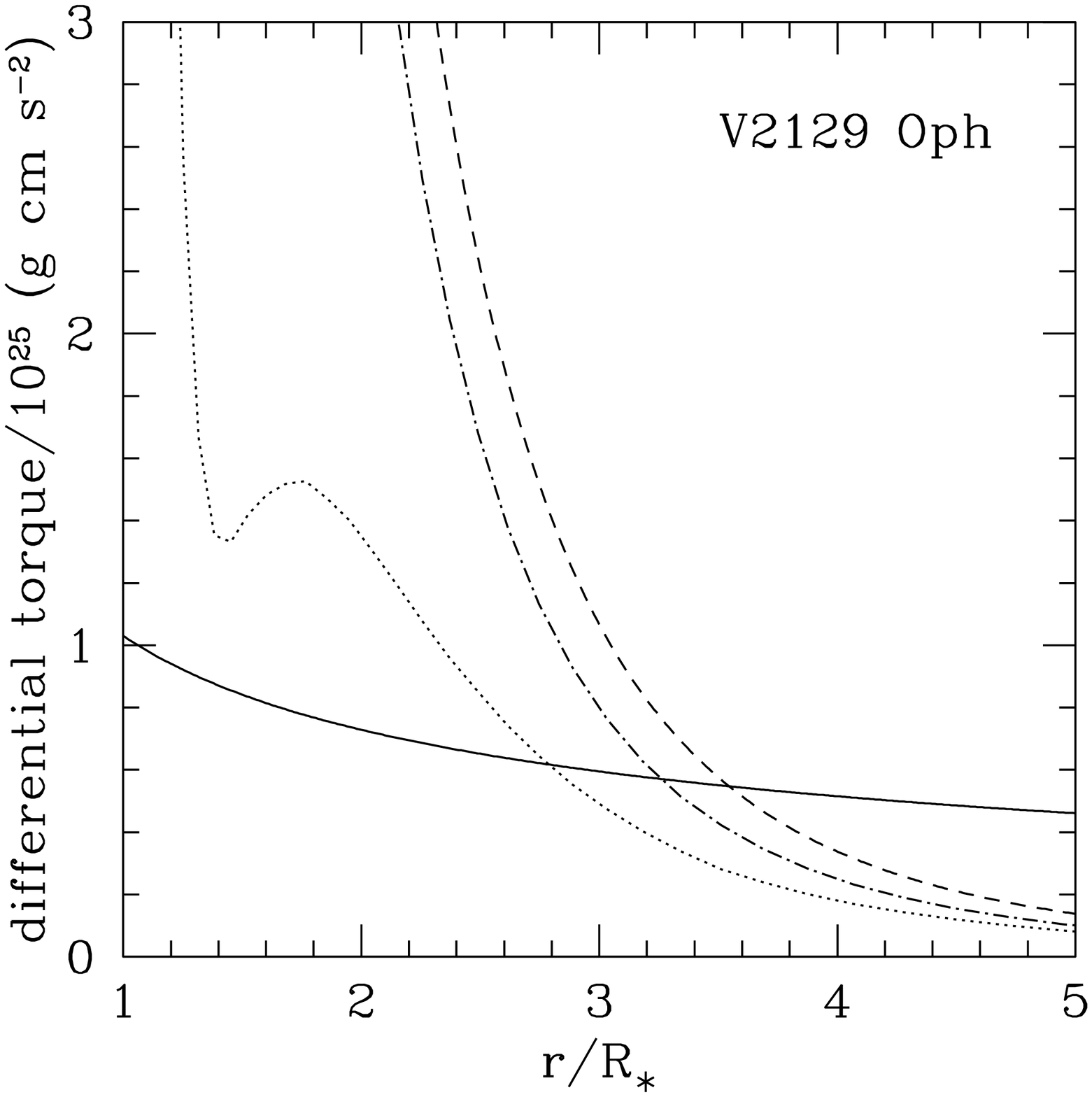,width=80mm}
                        }  \\
        \end{tabular}
        \caption[]{The variation along the equatorial plane of the differential magnetic/viscous torque [i.e. the LHS and RHS 
                   of equation (\ref{trunc})] for BP~Tau (left panel) 
                   and V2129~Oph (right panel).  The solid line represents the RHS of (\ref{trunc}), with the LHS for an aligned
                   dipole field (dashed line), a tilted dipole field with a source surface (dash-dot line) and the 
                   field extrapolation of V2129~Oph/BP~Tau (dotted line).  In both cases the disc is truncated well within
                   the corotation radius of $6.7\,R_{\ast}$ (V2129~Oph) and $7.4\,R_{\ast}$ (BP~Tau), although the exact location of 
                   $R_t$ is very sensitive to the assumed accretion rate.}
        \label{rt}
\end{figure*}

\citet{bes08} provide an overview of the various assumptions that have been used in the literature
to calculate the disc truncation radius.  Here we provide rough estimates of the inner disc radius for 
BP~Tau and V2129~Oph based on their extrapolated magnetic fields, and compare these values with those 
calculated by assuming both stars have dipole magnetic fields. 

We assume that the location of the inner disc is the radius at which the torque due to viscous 
processes in the disc is comparable to the magnetic torque due to the stellar magnetosphere.  This is 
equivalent to the approach taken previously by several authors, for example, \citet{cla95} and  
\citet{wan96}.  This torque depends on how much the disc is able to twist the
field lines of the stellar magnetosphere.  We approximate that this twist at the location of the inner
disc is of order 45\degr i.e. at the inner disc we assume that the perturbed toroidal
component is equal to the poloidal component ($B_{\phi} \sim B_z$).  This assumption should be valid
as long as the disc is truncated sufficiently within the corotation radius and assuming the disc is
strongly coupled to the stellar magnetic field \citep{mat05a}.  Equating the differential magnetic and 
viscous torques, and assuming that the poloidal field threading the disc is dominated by the vertical
component ($B_z \gg B_r$), gives
\begin{equation}
r^2B_z^2 = \frac{1}{2}\dot{M}\left (\frac{GM_{\ast}}{r} \right )^{1/2}.
\label{trunc}
\end{equation}
Note that this equation assumes a cylindrical corordinate system, however as we are restricting
the problem to the star's equatorial plane, the cylindrical radius and the spherical radius are
the same, and $B_z = B_{\theta}$.  Interestingly, the disc disruption radius found in the numerical
simulations of \citet{lon05} (see the discussion in their section 2) approximately conincides with
that derived through the use of equation (\ref{trunc}).  The solid line in Fig. \ref{rt} represents the variation in the 
RHS of this equation along the stars equatorial plane, using the stellar parameters
given in \S2.1.  The various dashed and dotted lines in Fig. \ref{rt} represent the LHS of this 
equation assuming different forms for the stellar magnetic field.    
The value of $r$ where the lines representing the LHS and RHS of equation (\ref{trunc}) cross, is then
the disc truncation radius $R_t$.  

For a dipole magnetic field the variation in $B_z$ along the equatorial plane can be written as
\begin{equation}
B_z = B_{\theta} = \frac{1}{2}B_{\ast, pole}\left( \frac{R_{\ast}}{r}\right)^3.
\label{bz_dipole}
\end{equation}
The dashed lines in Fig. \ref{rt}
represent such an aligned dipole field with a polar strength of $B_{\ast, pole}=1.2\,{\rm kG}$ for BP~Tau
and $0.35\,{\rm kG}$ for V2129~Oph, i.e. dipole fields with the same polar strength as the dipole 
components of the measured fields of both stars (see \S2.2).  With such an assumption for the stellar magnetic 
fields, we obtain a disc truncation radius of $\sim 5R_{\ast}$ for BP~Tau and $\sim 3.6R_{\ast}$ for 
V2129~Oph.  This is well within the corotation radius in both cases, justifying our assumption that the 
perturbed toroidal component of the field equals the poliodal component at the inner edge of the disc
($R_t \sim 0.7R_{co}$ for BP~Tau and $R_t \sim 0.5R_{co}$ for V2129~Oph).  However, for a better comparison
with the extrapolated fields we consider modified and tilted dipole fields, i.e. dipole
fields with a source surface, tilted by the same amount as the dipole component in both stars.
Once the dipole field been tilted, the magnetic field is no longer axisymmetric in the 
stars equatorial plane, and therefore $B_z$ is no longer uniform in azimuth at a fixed radius.  At each radius $r$ 
we therefore take the average of the square of $B_z$ for use in equation (\ref{trunc})\footnote{Once the dipole
has been tilted the field threading the disc has a small radial component, however for small dipole tilts
this remains much smaller than $B_z$.} (note that we follow the same
procedure for the extrapolated fields discussed below).
The dash-dot lines in Fig. \ref{rt} represent such tilted dipoles, with the disc truncation radius
being slightly closer to the stellar surface for such fields due to the non-uniformity of $B_z$ in azimuth.  
In Fig. \ref{rt} (and Figs. \ref{bratio_small} and \ref{bratio_big} below) we have chosen to set $R_S = 20\,R_{\ast}$ 
in order to minimise the effects of the source surface boundary condition upon the field structure close
to the accreting field regions.
  
We have already seen in \S3.1 that the larger scale field of BP~Tau is similar to a dipole, but that the field of 
V2129~Oph shows significant departures from such a simple field configuration.  We therefore expect that the disc 
truncation radius calculated using the extrapolated field of BP~Tau will be similar to that of a dipole 
while $R_t$ for V2129~Oph may be different.  The dotted lines in Fig. \ref{rt} demonstrate that this is indeed the case, 
with $R_t$ for BP~Tau closely matched to that of the modified dipole.  This is due to the strong dipole component
of BP~Tau's field, which contains 50\% of the magnetic energy \citep{don08a}, and is 4-times stronger than
the dipole component of V2129~Oph.  The calculated inner disc radius for V2129~Oph
is closer to the stellar surface than would be expected for a dipole field.   This is a reflection of the intrinsic 
field complexity of V2129~Oph, and the rapid drop off in field strength with height above the stellar surface for 
its magnetic field (see \S5.2 below).  When the magnetic field of the central star is particularly complex, as in the 
case of V2129~Oph, we find that the inner disc is truncated closer to the stellar surface than would be 
expected from dipole magnetic field models.  However, the use of equations such as 
(\ref{trunc}) in calculating disc truncation radii for individual stars should be treated with caution.  
Firstly, the values of $R_t$ determined here are very sensitive to the assumed mass accretion rates $\dot{M}$, which are
poorly constrained observationally.  For example, only a factor of 2 difference in our assumed accretion
rate for BP~Tau of $2.88 \times 10^{-8}\,{\rm M}_{\odot}{\rm yr}^{-1}$ would change the inner disc radius
from $4.8R_{\ast}$ to $3.9R_{\ast}$ or $5.9R_{\ast}$ depending on whether the accretion rate is increased or decreased
respectively.  Secondly, equation (\ref{trunc}) assumes that the field threading the disc is dominated by the 
vertical component $B_z$,
and although this is the case for the dipole-like large scale field of BP~Tau, it is not true for V2129~Oph
where the field has a significant radial component at the stellar equatorial plane.  Thus the shear within the disc
of V2129~Oph may generate a significant toroidal component, invalidating the use of equation (\ref{trunc}), and 
suggesting that our qualitative calculation of the disc truncation radius is too simplistic.
An MHD simulation using the extrapolated fields as starting points would be welcome here, in 
order to calculate inner disc radii more quantitatively.  However, our conclusion that the inner disc should
be truncated at, or within, the location obtained when considered a dipole magnetic field should remain valid.  


\subsection{Structure of the accreting field}
In the previous sections we compared the structure of the magnetic fields of V2129~Oph and BP~Tau to 
that of a dipole.  In this section we select out only those field lines which would be able to 
interact with the accretion disc and carry material on to the star.  In order
to select such field lines we follow the algorithm discussed by \citet{gre06a,gre07}, which we only 
summarise here.  

We consider a thin accretion disc and for V2129~Oph assume that the disc normal is 
parallel to the stellar rotation axis, such that the disc mid-plane is aligned with the stars equatorial
plane.  By extrapolating the field of BP~Tau, \citet{don08a} argued that 
a flat disc model could not explain the observed location of accretion hotspots.  For material to 
accrete into the observed hotspot locations, either the disc must be tilted relative to the stellar
rotation axis (i.e. aligned with the stellar magnetic equator - similar to the scenario 
found during the 3D MHD simulations of \citealt{rom03}), or the inner disc must be warped.    
Radio CO maps of BP~Tau's disc suggest a disc inclination of 30\degr offset from the stellar inclination of 
45\degr \citep*{sim00}, although both values are rather uncertain.  Analytic arguments suggest that a large scale 
tilted magnetosphere (as possessed by BP~Tau) will warp the inner disc, allowing material to accrete on to 
latitudes which are inaccessible if accretion were to proceed from a flat, wedge-shaped, disc \citep{ter00}.
Furthermore, the 3D MHD simulations of \citet{rom03}, suggest that complicated inner disc warps may arise
due to the interaction of the magnetosphere with the disc.  There is also strong evidence for the stellar
field warping the disc in at least one other star, AA~Tau (e.g. \citealt{bou07}), which is inclined
such that our line-of-sight to the star looks through the inner disc.  
Given the smaller inclination of BP~Tau ($i = 45\degr$), however, photometric eclipses due to a dense inner disc warp crossing our 
line-of-sight are not expected.  For BP~Tau we therefore assume that the disc is slightly tilted, or equivalently 
that the inner disc has been warped by the stellar magnetosphere\footnote{We note, however, that the 
differences in the accreting field structure shown in Figs. \ref{bratio_small} and \ref{bratio_big} for BP~Tau
are almost negligible when considering a flat or slightly warped disc.}.  An obvious question to ask, however, is
why did \citet{don07} find that a flat disc model was sufficient to explain the observed hotspot locations on V2129~Oph,
but \citet{don08a} find that a small inner disc warp was required for BP~Tau?  This is likely related to the strength
of the dipole component on both stars, which is almost 4-times stronger on BP~Tau than V2129~Oph.  At
the inner disc, at a distance of a few stellar radii, the dipole component is likely dominant, with the 
strength of the higher order field components dropping faster with height above the star.  The larger
scale magnetosphere of BP~Tau is therefore more likely to distort the circumstellar disc.  Indeed to explain the 
hotspot locations on BP~Tau the disc had to be tilted in the same direction as the dipole moment.  Tilting the disc
in a similar way for V2129~Oph makes little difference to the hotspot location, suggesting that it is the 
strength of the dipole that is affecting the inner disc.  Further numerical 
simulations of how magnetic fields with a realistically complexity can distort the structure of the disc will be 
required in future to confirm this, and are currently being undertaken.    

\begin{figure*}
        \def\subfigtopskip{4pt}
        \def\subfigbottomskip{4pt}
        \def\subfigcapskip{2pt}
        \centering
        \begin{tabular}{cc}
                \subfigure{
                        \label{bratio_big_bptau}
                        \psfig{figure=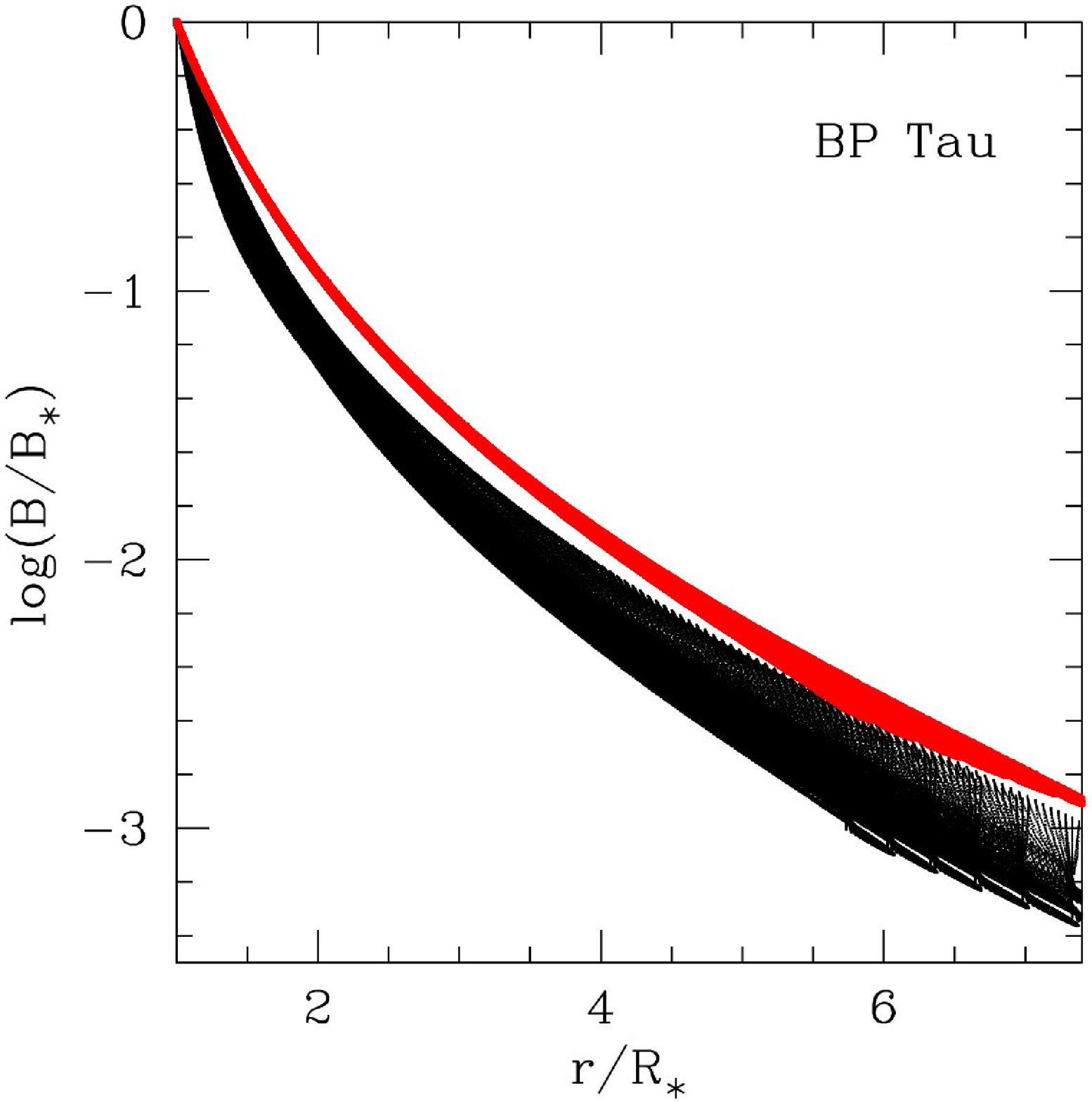,width=80mm}
                        } &
                \subfigure{
                        \label{bratio_big_v2129oph}
                        \psfig{figure=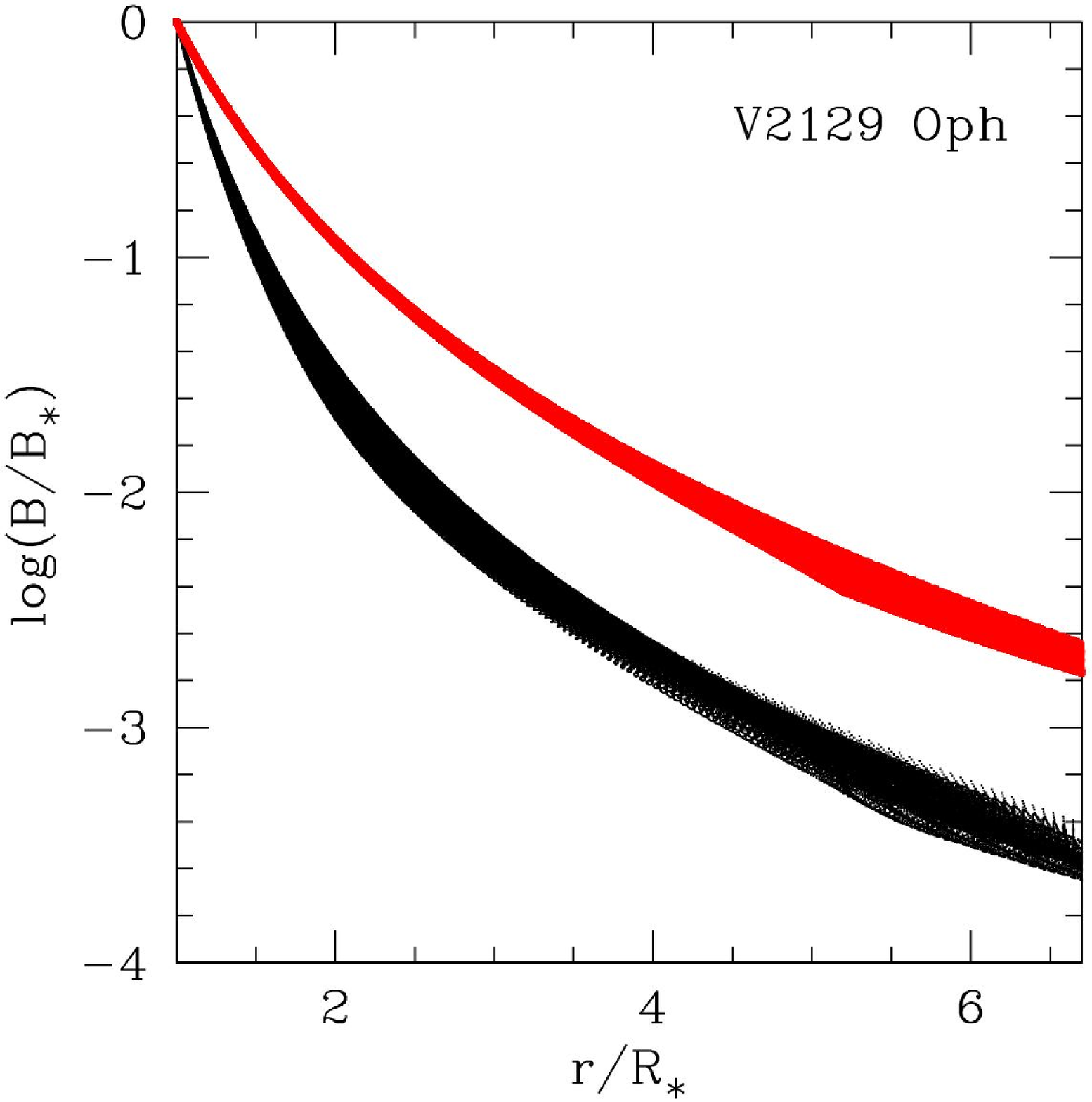,width=80mm}
                        }  \\
        \end{tabular}
        \caption[]{The drop in field strength with height above the surface of BP~Tau (left panel) and V2129~Oph (right panel) along 
                   the accreting field lines (black points).  Also shown for comparison is a dipole field (red points).  The dipole has been 
                   tilted by 10\degr (30\degr) for comparison with the field of BP~Tau (V2129~Oph).  Accretion is assumed to occur from
                   the corotation radius to $\sim 0.75\,R_{co}$.}
        \label{bratio_big}
\end{figure*} 

\begin{figure*}
        \def\subfigtopskip{4pt}
        \def\subfigbottomskip{4pt}
        \def\subfigcapskip{2pt}
        \centering
        \begin{tabular}{cc}
                \subfigure{
                        \label{bratio_small_bptau}
                        \psfig{figure=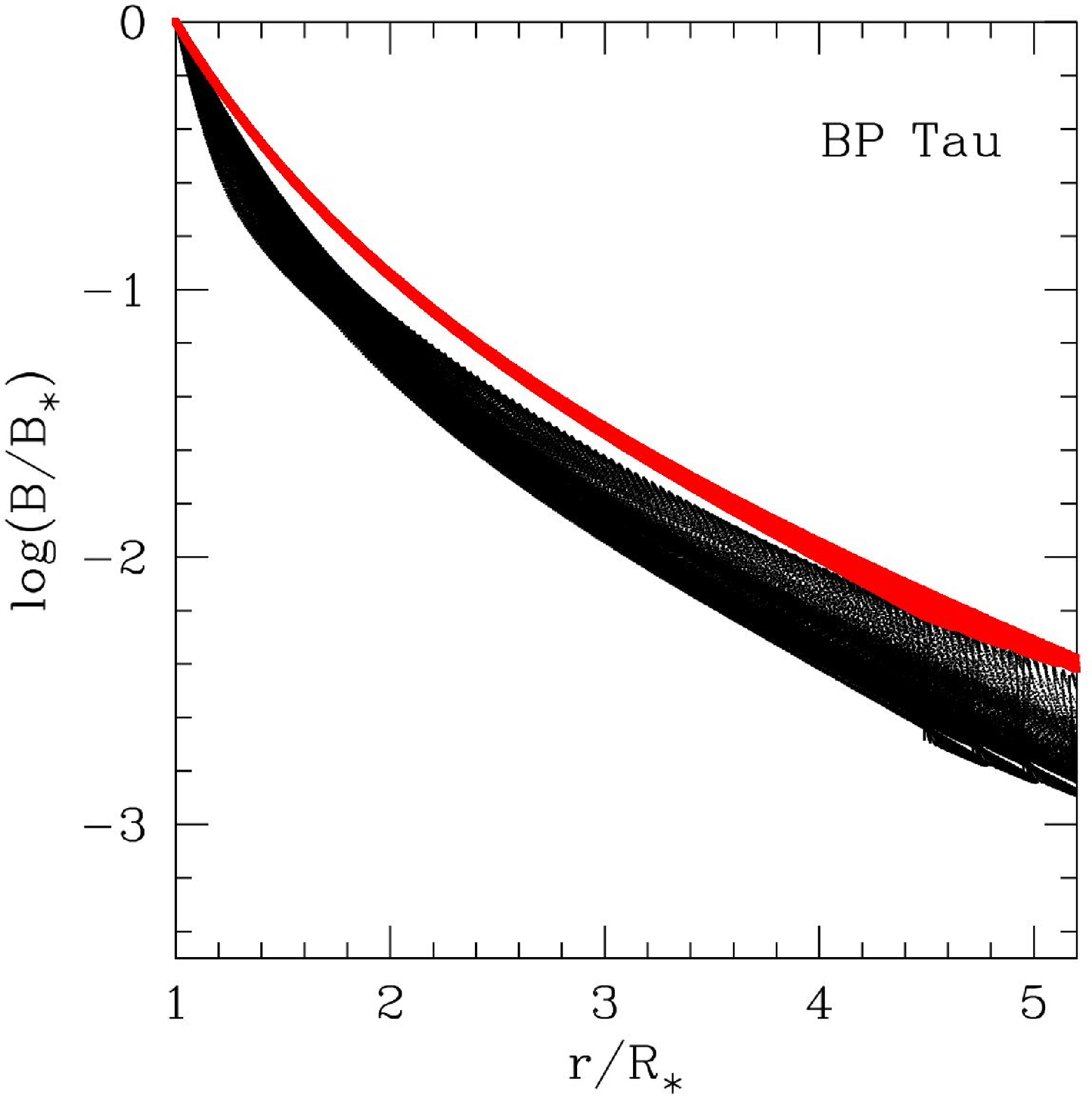,width=80mm}
                        } &
                \subfigure{
                        \label{bratio_small_v2129oph}
                        \psfig{figure=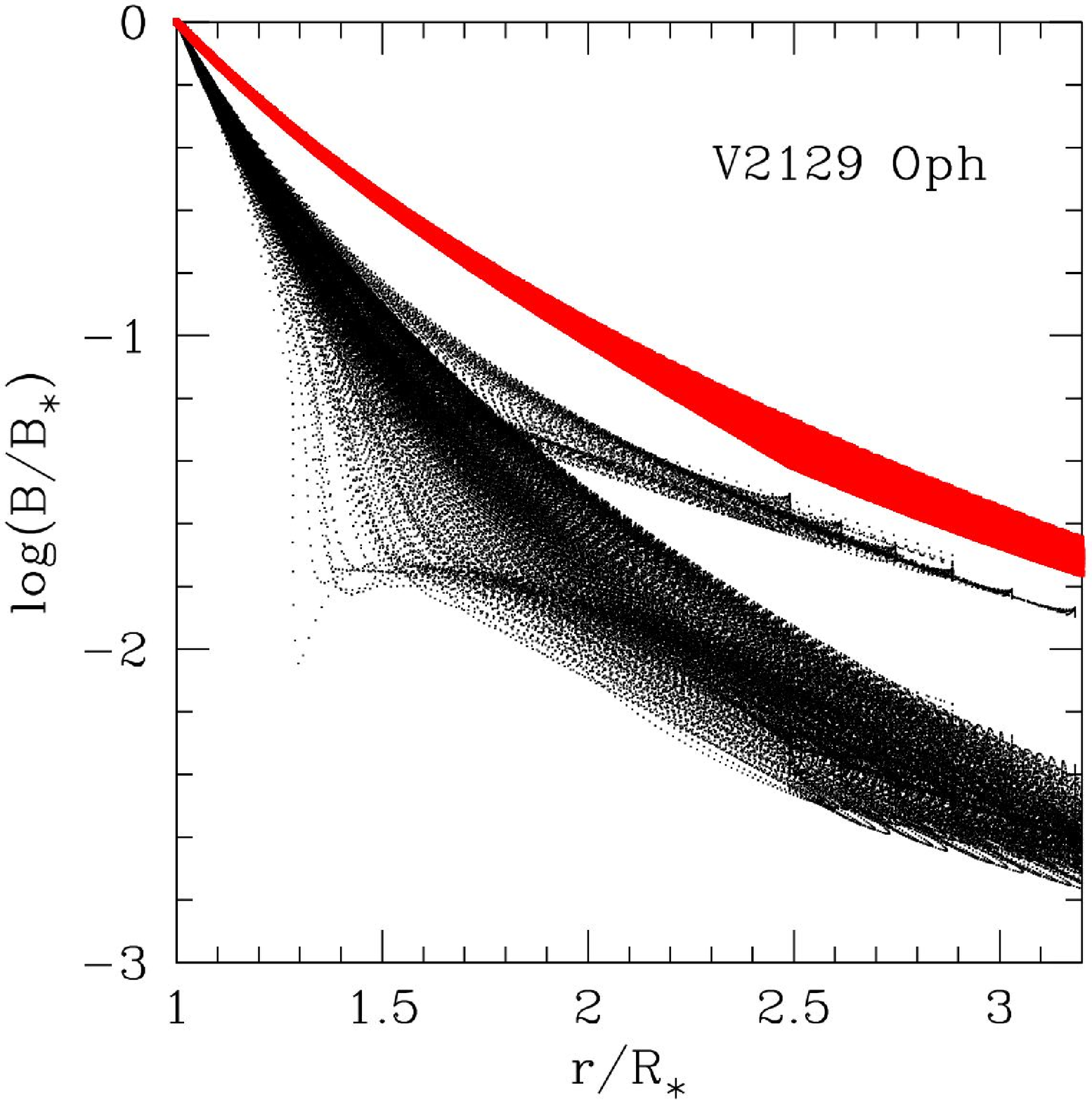,width=80mm}
                        }  \\
        \end{tabular}
        \caption[]{The same as Fig. \ref{bratio_big}, but assuming that accretion is occuring from a range of radii about the disc
                  truncation radius, as calculated in \S5.1 (see Fig. \ref{rt}).  The structure of the accreting field is shown
                  for BP~Tau (left hand panel) and V2129~Oph (right hand panel).}
        \label{bratio_small}
\end{figure*}

In order to compare the structure of the complex fields of V2129~Oph and BP~Tau with that
of a dipole we consider how fast the field strength drops with height above the stellar surface.
We therefore compare how the ratio $B(r)/B_{\ast}$ varies with $r$, where $B(r)$ is the strength
of the field at some position $r$ along the path of a field line and $B_{\ast}$ the field strength
at the field line footpoint.  $r$ is the spherical radius measured from the centre of the star.  
Figs. \ref{bratio_big} and \ref{bratio_small} show such plots for both V2129~Oph and BP~Tau.
All $B(r)/B_{\ast}$ values are plotted from the disc mid-plane to the field line footpoint at the stellar surface.
Accretion is assumed to occur along all field lines threading the mid-plane of the disc across a range of 
radii.  In Fig. \ref{bratio_big} we assume that accretion is occuring
from the corotation radius, down to 0.75$\,R_{co}$ ($\sim 5 - 6.7\,R_{\ast}$ for V2129~Oph and 
$\sim 5.5 - 7.4\,R_{\ast}$ for BP~Tau), whereas in Fig. \ref{bratio_small} we assume that  
accretion is occuring from a range of radii about the disc truncation radius calculated in \S5.1.  
For V2129~Oph, where the disc truncation radius was calculated to be $2.8\,R_{\ast}$ (see Fig. \ref{rt}), 
the accreting field is assumed to span the range of radii $\sim 2.4 - 3.2\,R_{\ast}$; for BP~Tau, with a disc
truncation radius of $\sim 4.8\,R_{\ast}$, accretion is assumed to occur from $\sim 4.4 - 5.2\,R_{\ast}$.
In practise little difference is found in the structure of the accreting field when these values are varied. 
Figs. \ref{bratio_big} and \ref{bratio_small} also show the behaviour of a dipole magnetic
field, tilted by 10\degr for comparison with the BP~Tau data (whose dipole component is tilted by 10\degr 
relative to the stellar rotation axis), and by 30\degr for comparison with the V2129~Oph data. 

Plots such as Figs. \ref{bratio_big} and \ref{bratio_small} allow a comparison of how the field complexity 
varies with height above the stellar surface.  We note, however, that we do not account for how the
structure of the magnetic field will evolve in time due to the interaction with the disc.  However, this
will be less of an issue if accretion proceeds from close to corotation, where the differential
rotation of field line footpoints anchored in the disc and on the stellar surface is much less than 
if the disc is truncated well within corotation.  As the strength of the dipole component decreases with height
more slowly than the higher order field components, the behaviour of the field on the largest scales 
should be somewhat simpler than that of the surface field.  This is the case for the magnetic field of 
V2129~Oph where the field behaves more like a dipole above $\sim 2.5R_{\ast}$ (see Fig. \ref{bratio_big}, 
right panel), after an initial fast drop within the strong complex field region close to the stellar surface.  
For BP~Tau the field is generally more dipolar, even closer to the stellar surface 
(see Fig. \ref{bratio_big}, left panel), reflecting the much stronger dipole component of it's magnetic field, 
found to be $1.2\,{\rm kG}$ and containing about $50\%$ of the magnetic energy \citep{don08a}.  For V2129~Oph 
if we assume that accretion occurs from around the disc truncation radius as calculated in the previous section,
see Fig. \ref{bratio_small}, we find that the structure of the accreting field is much more complex than
a simple dipole.  \citet{don07} found that the bulk of accretion in the visible hemisphere of V2129~Oph occured
into a single magnetic spot at high latitude.  Using the algorithm discussed in \citet{gre07} and \citet{jar08} we calculate
that about 45\% of the mass supplied by the disc would accrete into the high latitude northern hemisphere 
spot on V2129~Oph, with a similiar amount to an unseen southern hemisphere spot, and less than 10\% to smaller
equatorial spots.  Thus, provided that the source surface is set to a large enough radius 
to allow the longest field lines to connect to the high latitude spot, the bulk of the accreting material 
accretes into the high latitude spots, even if the disc is truncated within the corotation radius.  


\section{Conclusions and Discussion}
By extrapolating the fields of two T Tauri stars from observationally derived surface magnetograms
we have compared the resulting magnetic field topologies with a simple dipole.  We
have found that although the surface magnetic fields of both stars are particularly complex, 
the larger scale field is simpler and more well ordered.  
This is consistent with previously published spectropolarimetric studies of accreting
T Tauri stars, which suggested that bundles of accreting field lines connect to the stellar surface
in single polarity regions, and were therefore likely well-ordered (e.g. \citealt{val04}).
The larger scale field of BP~Tau in particular is closely matched to a dipole magnetic field, although
V2129~Oph still shows departures from a dipole field even on the largest scales.  However, for both stars, 
the footpoint separation at the stellar surface of the largest scale field lines is greater
than for a purely dipolar magnetic field.  In other words as the largest scale field lines arrive
at the stellar surface, their structure is influenced by the much stronger field regions.  
The effect of this is that the largest scale field connects to the star at higher latitudes than 
would be expected when using a dipole magnetic field model.  Thus by considering complex magnetic
field topologies we find that the fractional open flux through the stellar surface is less than would
be expected from dipole field models.  The reduction in the fractional open flux is mainly attributable
to the larger closed flux for the complex fields, which arises due to the numerous small magnetic loops
close to the stellar surface.  For both V2129~Oph and BP~Tau there is less total open flux than for
a dipole field, however, the difference is small enough (see Fig. \ref{fluxes}) that 
there is unlikely to be any significant implications for the amount of angular momentum that
can be carried away from such systems by a stellar wind \citep{mat05b,mat08a,mat08b}.  However, more 
spectropolarimetric data for many different T Tauri stars will be required to confirm this. 

If the extent of stellar magnetospheres are limited to well within corotation then accretion
may proceed along the more complex regions of the stellar surface field.  The result of this would be 
the formation of many low-latitude accretion hotspots \citep{gre05,gre06a}.  There is at least one
star which shows evidence for such equatorial spots, GQ~Lup \citep{bro07}.  However, for
V2129~Oph and BP~Tau the spectroscopic data of \citet{don07,don08a} clearly showed that
the bulk of accreting material is carried into high latitude regions of the stellar surface.  
In order to explain such high latitude hotspots, \citet{don07} and \citet{jar08} demonstrated that for V2129~Oph 
the source surface must be set to at least 6.0$\,{R_{\ast}}$, while for BP~Tau the stellar magnetosphere must
extend to at least 4$\,{R_{\ast}}$ \citep{don08a}.  Selecting the location of the source surface
sets the structure of the stellar magnetic field, although the inner disc may still be truncated closer
to the stellar surface.  Using the assumption that inner disc is located at the point where the torque
due to viscous processes in the disc is comparable to the magnetic torque due to the stellar magnetosphere
we estimated the disc truncation radius using the extrapolated fields.  For BP~Tau we found that the 
inner disc would be truncated close to the radius predicted for a dipole field, whereas for V2129~Oph, the inner
disc is likely to be closer to the star.
However, with large uncertainties in the assumed mass accretion rate (which influences the torque due to viscous 
processes in the disc), and with the static field structures considered here, such calculations are only of qualitative 
value.  Further 3D MHD simulations, such as those already performed by, for example \citet{lon08}, are required in order 
to model how magnetic fields with an observed degree of complexity can disrupt and influence the structure of planet 
forming discs, and will represent a major future development in this field.     

Although the radius of the source surface is essentially a free parameter of our model, the observed
hotspot locations provide constraints.  Once the source surface has been selected, however, this then also 
sets the structure of the X-ray emitting surface field, allowing predictions to be made
regarding the stellar X-ray emission properties \citep{jar08}.  Such predictions will allow direct 
testing of our field extrapolation model with future independently obtained X-ray observations.
A further test of the model will be the simulation of accretion related emission line profiles, and are
currently being undertaken. 

V2129~Oph, despite its young age, is massive enough to have developed a radiative core.  In contrast to this
BP~Tau is likely to be completely convective.  Although the field of BP~Tau is more complex than a dipole, it is 
much simpler than that of V2129~Oph.  The differences in the field structure of both stars likely reflect
the different internal structure and the process by which their magnetic fields are generated and 
maintained.  For stars with radiative cores the magnetic field is likely generated in the 
shear layer between the core and the outer convective envelope.  In the absence of such a shear layer, 
it is difficult to explain how fully convective stars can generate and maintain
large-scale almost axisymmetric kilo-Gauss fields, although recent theoretical models suggest that this
is possible \citep{bro08}.  With only three magnetic surface maps available to date, V2129~Oph and two of BP~Tau 
\citep{don07,don08a}, it is not possible to determine whether this is a general feature of all T Tauri stars, however it 
is similar to what it found for low mass main sequence stars \citep{don08b}. 
More magnetic surface maps of T Tauri stars with varying stellar parameters are now required to test these
ideas fully, and are an essential requirement to advance our understanding of stellar magnetism on 
the pre-main sequence.

\section*{Acknowledgements}
The authors wish to thank the anonymous referee for comments which have helped to clarify 
a number of detailed points within this paper.

\bibliographystyle{mn2e}
\bibliography{gregory}


\bsp

\label{lastpage}

\end{document}